\documentclass[aps,prd,nofootinbib,showkeys,preprint,floatfix]{revtex4}
\usepackage{amssymb}
\usepackage{amsmath}
\usepackage{amsfonts}
\usepackage{graphicx}
\usepackage{color}
\usepackage{xspace}
\usepackage{ulem}
\usepackage{lscape}
\usepackage{ wasysym }
\usepackage{multirow,rotating}
\usepackage{mathrsfs} % \mathscr

\def\gsim{\raise0.3ex\hbox{$\;>$\kern-0.75em\raise-1.1ex\hbox{$\sim\;$}}}
\def\lsim{\raise0.3ex\hbox{$\;<$\kern-0.75em\raise-1.1ex\hbox{$\sim\;$}}}
\def\esl{E \hspace{-0.6em}/\;\:}

\newcommand{\ba}[1]{\begin{eqnarray} \label{(#1)}}
\newcommand{\ea}{\end{eqnarray}}

\allowdisplaybreaks[4]

\renewcommand{\vec}[1]{\boldsymbol{{\mathrm #1}}}

\newcommand{\AddrAHEP}{Instituto de F\'{\i}sica Corpuscular --
  CSIC - Universitat de Val{\`e}ncia, 
  Parc Cient\'{\i}fic-UV, c/ Catedr\'atico Jos\'e Beltr\'an, 2,
  E-46980 Paterna (Val{\`e}ncia), Spain}

\newcommand{\AddrLaSerena}{
{\it Departamento de F\'{i}sica, Facultad de Ciencias, Universidad de La
Serena, Avenida Cisternas 1200, La Serena, Chile,
}
}

\newcommand{\AddrSAPHIR}{
  {\it Millennium Institute for Subatomic Physics at the High Energy Frontier
    (SAPHIR), Fern\'{a}ndez Concha 700, Santiago, Chile
}
}

\begin{document}

\title{Invisible neutron decay and light BSM particles}

\author{\large J.C. Helo}\email{jchelo@userena.cl}\affiliation{\AddrLaSerena
  \\\AddrSAPHIR}

\author{\large M. Hirsch}\email{mahirsch@ific.uv.es}
\affiliation{\AddrAHEP\vspace*{1cm} }

\author{\large T. Ota}\email{toshihiko.ota@userena.cl}
 \affiliation{\AddrLaSerena\\\AddrSAPHIR}


%\pacs{14.60.Pq, 12.60.Jv, 14.80.Cp}

%%%%%%%%%%%%%%%%%%%%%%%%%%%%%%%%%%%%%%%%%%%%%%%%%%%%%%%%%%%%%%%%%%%%%%
\begin{abstract}

\medskip
In Standard Model Effective Field Theory (SMEFT) invisible neutron decay arises from $d=12$ operators. Adding new, light particles to the field content of the SM, such as right-handed neutrinos, allows to construct operators for invisible neutron decay at much lower dimensions. Observing invisible neutron decay, if nucleon decays with charged leptons remain absent, would therefore point towards the existence of new neutral degrees of freedom. Here, we discuss four cases: (i) Adding right-handed neutrinos to the SM; (ii) a right-handed neutrino and an axion-like particle; (iii) a right-handed neutrino and a (nearly) massless singlet scalar; and (iv) a right-handed neutrino and a light $Z'$. We give the general tree-level decomposition for the resulting $d=(7-9)$ operators for invisible neutron decay.  We also briefly discuss LHC searches related to the exotic states found in these UV completions.

\end{abstract}

%%%%%%%%%%%%%%%%%%%%%%%%%%%%%%%%%%%%%%%%%%%%%%%%%%%%%%%%%%%%%%%%%%%%%%
\maketitle

%%%%%%%%%%%%%%%%%%%%%%%%%%%%%%%%%%%%%%%%%%%%%%%%%%%%%%%%%%%%%%%%%%%%%%
\tableofcontents

%%%%%%%%%%%%%%%%%%%%%%%%%%%%%%%%%%%%%%%%%%%%%%%%%%%%%%%%%%%%%%%%%%%%%%
% !TEX root = ../InvNDecay.tex
\section{Introduction}

Searches for nucleon decays by Super-Kamiokande have pushed limits to
$10^{34}$ yr and above for several decay channels.  
The best known examples are probably $p \to \ell^+ \pi^0$
($\ell=e,\mu$) \cite{Super-Kamiokande:2020wjk} and $n \to \bar{\nu} K^0$
\cite{Super-Kamiokande:2025ibz}, which are motivated by Grand
Unified Theories (GUTs).
Less known are decays to three charged leptons, where Super-Kamiokande 
also gives limits in excess of $10^{34}$ yr 
now \cite{Super-Kamiokande:2020tor}.  
Limits on many other channels, however, are less stringent, sometimes 
by orders of magnitude, 
see \cite{Takhistov:2016eqm, ParticleDataGroup:2024cfk}. 
On the other hand, nucleon decay searches form an important part of the
agenda of next generation underground detectors such as JUNO
\cite{JUNO:2015zny}, Hyper-Kamiokande \cite{Hyper-Kamiokande:2018ofw}
or DUNE \cite{DUNE:2020lwj} and significant improvements in
sensitivity on many nucleon decay modes are expected from these
experiments in the foreseeable future.

Traditionally, searches for proton decay were motivated by the
predictions from GUTs, such as the classical $SU(5)$
\cite{Georgi:1974sy}, for a review see for example
\cite{Nath:2006ut}. 
More recently, however, many theoretical papers have taken a more
agnostic view and studied nucleon decays from the point of view of
effective field theory (EFT) \cite{deGouvea:2014lva,
  Kobach:2016ami,Hambye:2017qix, Fonseca:2018ehk,Helo:2018bgb,
  Helo:2019yqp,Heeck:2019kgr, Beneito:2023xbk,Fridell:2023tpb,
  Gargalionis:2024nij,Heeck:2024jei,Li:2024liy,
  Li:2025slp,IBeneito:2025nby,Heeck:2025btc,Heeck:2025uwh,Fan:2025xhi}.\footnote{%
Here we also cite some pioneering works on nucleon decays in
SMEFT~\cite{Weinberg:1979sa,Wilczek:1979hc,Weinberg:1980bf,
  Weldon:1980gi,Abbott:1980zj,Claudson:1981gh} and the studies on the
decompositions of the relevant effective
operators~\cite{Bowes:1996xy,Kovalenko:2002eh,Arnold:2012sd,
  Assad:2017iib,deBlas:2017xtg,Li:2023cwy}.}
In Standard Model Effective Field Theory (SMEFT), one can
classify different decay modes according to the dimensionality of the
corresponding operators. For example, two-body decays with
$\Delta(B+L)=2 $ are described by $d=6$ operators, $\Delta(B-L)=2$ are
$d=7$, {\em genuine} three-body decays need at least $d=9$ operators
\cite{Heeck:2019kgr}. 

In the current paper, we are interested in invisible neutron decay.
Considering only SM particles, the only possibility for a neutron to
decay into ``nothing'' is the channel $n \to 3 \nu$.\footnote{Here
both $\nu$ and $\bar{\nu}$ are mentioned as $\nu$.} 
In SMEFT one has to go up to $d=12$, to find for the first time an
operator capable of generating invisible neutron decay, without the
corresponding decays to charged leptons, as already realized by
Weinberg more than 40 years ago \cite{Weinberg:1980bf}.  Weinberg gave
as an example the operator \cite{Weinberg:1980bf}:
\begin{equation}\label{eq:W12}
  \mathscr{L}_{d=12}^{\rm W} = \frac{c_{12}^{\text{W}}}{\Lambda^8}
  \epsilon_{IJK}
  \epsilon^{ij}
  \epsilon^{kl}
  \epsilon^{mn}
  \left( \overline{u_{R}}^{I} L_i \right)
  \left( \overline{d_{R}}^{J} L_k \right)
  \left( \overline{d_{R}}^{K} L_m \right)
  H_{j}
  H_{l}
  H_{n},
\end{equation}
where $\Lambda$ is the typical scale of the origin of this effective
operator and $c_{12}^{\text{W}}$ is the Wilson coefficient.
In Eq.~\eqref{eq:W12} we have suppressed generation indices for
simplicity, $\epsilon_{IJK}$ and $\epsilon^{ij}$ are anti-symmetric
tensors for the contraction of $SU(3)_{c}$ and $SU(2)_{L}$ indices.
For this $d=12$ operator one can estimate the nucleon decay width for
the 3-body final state $n\to 3 \nu$ roughly as \cite{Heeck:2019kgr}
\begin{eqnarray}\label{eq:wdth12}
  \Gamma \sim \frac{\beta_h^2 m_N^{5} \langle H^{0} \rangle^6}
         {6144 \pi^3 \Lambda^{16}}
  & \Rightarrow & \frac{1}{9 \cdot 10^{29}\hskip1mm {\rm yr}}
  \Big(\frac{13.4\hskip1mm {\rm TeV}}{\Lambda}\Big)^{16} .
\end{eqnarray}
Here, $\beta_h \simeq 0.014$ GeV$^3$ is the nuclear matrix element
\cite{Aoki:2017puj}, $m_N$ the nucleon mass, $\langle H^{0} \rangle \simeq
174$ GeV the vacuum expectation value of the SM Higgs field, 
and $c_{12}^{\text{W}}$ is taken to be unity.

Since the operator in Eq.~\eqref{eq:W12} by itself will not
  generate final states with charged leptons at any measurable level,
  one might be tempted to assume that just having a low enough scale
  $\Lambda$ would be sufficient to create an observable rate for the
  invisible neutron decay. However, this statement neglects that there
  are lower dimensional operators that will generate neutron decays
  with charged lepton final states. SMEFT operators with three leptons
  have been discussed in \cite{Hambye:2017qix}. In SMEFT, the lowest
  dimensional operator with three leptons has $d=9$, one example is
  ${\cal O} \propto (QL)({\bar L}{\bar L})(d_Rd_R)$. Ref.
  \cite{Hambye:2017qix} lists a total of 16 operators of this type at
  $d=9$. Some of these $d=9$ operators can generate the decay $n \to 3
  \nu$. However, all the operators that can do so, always also
  generate decays to charged leptons with very similar rates.  Since
  nucleon decay searches are much more sensitive to final states with
  charged leptons \cite{Super-Kamiokande:2020tor} than to $n \to 3
  \nu$ \cite{ParticleDataGroup:2024cfk}, none of the $d=9$ operators
  can generate invisible neutron decay at observable rates. Moreover,
  if any of these $d=9$ operators is present in the theory, one expects
  it to dominate the nucleon decay rate by many orders of magnitude
  w.r.t the $d=12$ operator in Eq.~\eqref{eq:W12}.

To avoid this conclusion, one could argue that all $d=9$ operators
  are actually forbidden by some symmetry, such that the $d=12$
  operator in Eq.~\eqref{eq:W12} is the lowest dimension at which
  neutron decay is possible. However, this is not possible. While some
  specific symmetry could indeed eliminate all $d=9$ operators, at
  $d=10$ one finds an operator with three leptons
  \cite{Weinberg:1980bf}: ${\cal O} \propto d^c d^c d^c L L L
  H$.\footnote {This operator is non-zero only for two or more
  generations of down quarks {\it or} leptons.}  This operator induces
  the decay $n \to e^- \pi^+ \nu\nu$ \cite{Fonseca:2018ehk} and will
  dominate the decay rate, unless its Wilson coefficient is fine-tuned
  to be very close to zero. As we show in appendix \ref{app2} one cannot 
  forbid this operator by any symmetry, if the $d=12$ operator is
  to be allowed. Thus, within pure SMEFT, one expects that invisible
  neutron decay is always accompanied by neutron decays with charged
  lepton final states.

For invisible neutron decay the particle data booklet
\cite{ParticleDataGroup:2024cfk} cites an old Kamiokande result (from
1993) \cite{Kamiokande:1993ivj} ($4.9\cdot 10^{26}$ yr), but there are
more recent and stronger bounds set by 
SNO~\cite{SNO:2003lol} ($2 \cdot 10^{29}$), 
KamLAND~\cite{KamLAND:2005pen} ($5.8\cdot 10^{29}$ yr, 
cf. also \cite{Kamyshkov:2002wp}) 
and two analysis by the SNO+ collaboration \cite{SNO:2018ydj} 
($2.5\cdot 10^{29}$ yr) and \cite{SNO:2022trz} ($9.0\cdot 10^{29}$ yr). 
For the nearer and farther future sensitivity estimates have been given by
JUNO \cite{JUNO:2024pur} ($5 \cdot 10^{31}$ yr) and for THEIA (80kt),
that could ultimately reach $\sim 4 \cdot 10^{32}$ yr according to
their proposal~\cite{Theia:2019non}.

However, there exists a simple possibility to generate invisible
neutron decay with lower dimensional operators {\it and without
  generating sizeable rates into final states with charged leptons}:
We can extend the SM particle content with new light degrees of
freedom.
In this paper we discuss four different possibilities. 
First we add (i) (at least two) right-handed neutrinos $N_{R}$, also
sometimes called sterile neutrinos in the literature, cf.
\cite{Acero:2022wqg,Dasgupta:2021ies,Diaz:2019fwt,Boser:2019rta,
  Dentler:2018sju,Kopp:2013vaa,Abazajian:2012ys} for reviews and
current bounds, motivated from various neutrino
experiments~\cite{LSND:2001aii,MiniBooNE:2010idf,MiniBooNE:2020pnu,
  Elliott:2023cvh,Barinov:2021asz,Dentler:2017tkw,STEREO:2022nzk}.
In addition, the introduction of such light degrees of freedom is often
discussed in the context of cosmology, 
e.g., 
%it is known that the extra radiation component mitigates
%the Hubble tension~\cite{Allali:2024cji,DiValentino:2021izs,Seto:2021xua},
%and 
the stringent bound on neutrino masses from recent cosmological 
observations~\cite{DESI:2024mwx,DESI:2024hhd,DESI:2025zgx,DESI:2025ejh}
maybe be alleviated by the mixing between
active neutrinos and massless sterile 
neutrinos~\cite{Farzan:2015pca,Escudero:2022gez,Benso:2024qrg,Ota:2024byu}.
In $N_R$SMEFT, for a list of basis operators see
\cite{Bhattacharya:2015vja,Liao:2016qyd,Li:2021tsq}, invisible neutron
decay operators appear at $d=9$.
Second, we add (ii) a right-handed neutrino and an axion-like particle
(ALP), $a$. 
The axion is one of the best-motivated light new physics
fields~\cite{Weinberg:1977ma,Wilczek:1977pj}. It arises as a
consequence of the breaking of the Peccei-Quinn symmetry introduced to
solve the strong CP problem~\cite{Peccei:1977hh,Peccei:1977ur}.
For experimental searches and phenomenology of ALPs, cf. e.g.,
\cite{Graham:2015ouw,Brivio:2017ije,Bauer:2017ris,Irastorza:2018dyq,
  Bauer:2018uxu,Gavela:2019wzg,Merlo:2019anv,Alda:2025uwo} and for
cosmological aspects of ALPs, cf. e.g.,
\cite{Cadamuro:2011fd,Arias:2012az,Marsh:2015xka,Dror:2021nyr,
  Chadha-Day:2021szb,Adams:2022pbo}.
The basis operators of $a$SMEFT, the effective theory with the SM
particle content plus an ALP, are listed in \cite{Song:2023lxf,
  Song:2023jqm,Grojean:2023tsd}.
In $a N_R$SMEFT, i.e. SMEFT with $N_{R}$s and ALPs, invisible neutron
decay operators appear at $d=8$.\footnote{Basis operators with a
given particle content can be constructed up to a given mass dimension
$d$ by \texttt{AutoEFT}~\cite{Harlander:2023ozs} or also
\texttt{Sym2Int} \cite{Fonseca:2017lem} }

Finally, we add a right-handed neutrino and (iii) a (nearly) 
massless scalar, $\phi$, or (iv) a light vector boson, $Z'$.
A ultra-light scalar is an attractive candidate of dark matter, 
aka ``fuzzy'' dark matter~\cite{Hui:2016ltb}. 
%which may alter the neutrino oscillation probability in a particular
%way~\cite{Berlin:2016woy,Brdar:2017kbt,Liao:2018byh,Capozzi:2018bps,Ge:2018uhz}.
%
For experimental searches and cosmologcal effects of the light vector
boson, aka dark photon, cf. e.g., \cite{Agrawal:2018vin,McDermott:2019lch,
  Fabbrichesi:2020wbt,Caputo:2021eaa}.
The basis operators of $\phi$SMEFT, SMEFT with a SM singlet scalar,
and $Z'$SMEFT, SMEFT with a dark photon, 
are listed in \cite{Song:2023lxf,Song:2023jqm}.\footnote{%
Later we will use the symbol $S$ for the scalars that mediate the
effective operators, and therefore, to avoid the confusion,
we use $\phi$ for the light scalar field in the final state of the invisible
neutron decay process, although the effective theory
with $\phi$ is called sometimes $s$SMEFT in the literature.
} 
The invisible decay of neutrons is generated from 
$d=7$ operators in both $\phi N_R$SMEFT (SMEFT with $\phi$ and $N_{R}$)
and $Z'N_R$SMEFT (SMEFT with $Z'$ and $N_{R}$)~\cite{Liang:2023yta}.

In this paper, we will give the operators for invisible neutron decay
for all four cases, discussed above, estimate the neutron decay
half-lives and give the complete list of the possible UV completions
for these operators at tree-level. For the different BSM particles
found in these UV models we will also briefly discuss current LHC
limits.

This introduction would not be complete without mentioning that
invisible neutron decay has been discussed in some earlier papers. 
For example, in \cite{Dvali:1999hn} the authors speculated about 
``decays to nothing'' in models with extra time dimensions. 
The authors of \cite{Mohapatra:2002ug} considered universal extra 
dimensions and an extended gauge group, to have the neutron 
decay dominantly into $n\to \nu \bar{\nu}_{s} \bar{\nu}_{s}$, where $\nu_{s}$ 
is some sterile neutrino.
The effective operators relevant 
to $n \rightarrow \bar{\nu}_{s} \bar{\nu}'_{s} \bar{\nu}''_{s}$
are listed in \cite{Girmohanta:2019fsx}
in the context of an extra dimension model.
In \cite{Girmohanta:2020eav} invisible neutron decay was studied in a
left-right symmetric model with large extra dimensions.  
Some recent works have considered also the possibility of 
a ``dark'' (invisible) neutron decay~\cite{Fornal:2018eol,Barducci:2018rlx,
  Cline:2018ami,Berezhiani:2018udo,Fornal:2020gto} in relation to the
neutron life-time anomaly \cite{Czarnecki:2018okw}. 
And, finally, a particular UV model for invisible di-neutron decay 
to two singlet fermions has been studied in ~\cite{Hao:2022xyj}, 
where the authors discussed also possible constraints from neutron stars.

The rest of this paper is organized as follows. In section 2 we will
introduce the different model variants in a bit more detail, discuss
some generalities, define the operators and estimate the neutron decay
half-lives for the different cases.  In section 3 we will then
decompose the operators to find the possible UV models at
tree-level. Section 4 discusses a number of different LHC searches and
how they can be used or reinterpreted to derive limits on the BSM
states found in section 3. We then close with a short discussion and
summary. Some tables for the decompositions will be relegated to the
appendix for the sake of keeping the main text more compact.

% !TEX root = ../InvNDecay.tex
\section{Setup: Models and effective operators \label{sect:ops}}

In this section we will describe the theoretical setup. We consider
four different extensions of the SM particle content. For each
case, we briefly describe the model. We then define the operators
relevant for invisible neutron decay. We give an estimate of the
neutron decay half-life in each case. Finally, in an additional
subsection we discuss a possible caveat. The decomposition of the
operators, defined here, is discussed in section 3.

\subsection{Variant I: Right-handed neutrinos \label{subsect:NR}}

In the standard model neutrinos are massless. While there are many
extensions of the SM that can explain neutrino masses, as observed
in oscilllation experiments, maybe the simplest possibility is to
add right-handed neutrinos to the SM particle content.

Massive neutrinos can be either Dirac or Majorana particles. In the
classical type-I seesaw model \cite{Minkowski:1977sc,Yanagida:1979as,
  Mohapatra:1979ia,Schechter:1980gr} right-handed neutrinos have
a Majorana mass term and the Lagrangian can be written as:
\begin{equation}\label{eq:Yuk}
\mathscr{L}_{Y} 
 = 
 (Y_{\nu}^{*})_{\alpha j} 
 \overline{N_{R j}} L_\alpha \cdot H
 + 
 \frac{1}{2}(M_{M})_{jj}
 \overline{{N_{R j}}^{c}} N_{R j} 
 +\hskip2mm {\rm h.c.}
\end{equation}
Note that $M_M$ can be taken diagonal without loss of generality. 
Majorana neutrinos, on the other hand, violate lepton number by
two units. Whether such a $\Delta L=2$ term is allowed or not,
will depend on the ultra-violet completion of the operators
defined below. We note, however, that the invisible neutron decay
$n \to 3 \bar{N}$ violates $L$ by three units, thus, 
Majorana masses need not to be allowed in general and neutrinos
could be Dirac particles, as far as invisible neutron decay is
concerned.

For the fit to current neutrino data, at least two copies of $N_R$ are
needed and both Dirac and Majorana neutrinos can explain existing
data. We will consider three copies of $N_R$, to mirror the three SM
generations, but doing an exact fit to oscillation data is irrelevant
for the neutron decay we are interested in. The only condition for the
decay to occur is that the right-handed neutrino masses are smaller
than $m_n/3$.

Two operators can be constructed at level $d=9$, that contribute
to invisible neutron decay:
\begin{eqnarray}
 \label{eq:opsNR1}
  {\cal O}_1^N & = & 
  \frac{c_{\alpha\beta\gamma ijk}^{(1)}}{\Lambda^5}
  \left(\overline{{u_{R \alpha}}^{c}} d_{R \beta} \right) 
  \left(\overline{{d_{R \gamma}}^{c}} N_{R i} \right)
  \left(\overline{{N_{R j}}^{c}} N_{R k} \right) , 
  \\ 
 \label{eq:opsNR2}
  {\cal O}_2^N & = &  
  \frac{c_{\alpha\beta\gamma ijk}^{(2)}}{\Lambda^5}
  \left( \overline{{Q_{\alpha}}^{c}} Q_{\beta} \right) 
  \left( \overline{{d_{R \gamma}}^{c}} N_{R i} \right)
  \left( \overline{{N_{R j}}^{c}} N_{R k} \right) .
\end{eqnarray}  
Here, $\alpha,\beta,\gamma$ are SM generation indices, while $i,j,k$
runs over the generation of $N_R$. To keep the expressions simpler, we
have suppressed $SU(3)_{c}$ (and $SU(2)_{L}$) indices. Here, and in all cases
discussed below the colour of the three quarks must be contracted with
a completely anti-symmetric tensor, $\epsilon_{IJK}$. It is
straightforward to show that both of these operators vanish in case
there is only one copy of $N_R$, we have checked this also with
\texttt{Sym2Int} \cite{Fonseca:2017lem}.

Assuming the mass of the right-handed neutrinos to be negligible and
one of the $c_{111ijk}^{(1/2)}=1$, we can estimate the partial half-life for the
invisible neutron decay as
\begin{align}
 \Gamma(n \rightarrow \bar{N}_{i} \bar{N}_{j} \bar{N}_{k})
 \simeq
 \frac{\beta_{h}^{2} m_{n}^{5}}{6144 \pi^{3} \Lambda^{10}}
 \simeq
 \frac{1}{9 \cdot 10^{29} \text{yr}}
 \left(
 \frac{180 \text{TeV}}{\Lambda}
 \right)^{10}.
 \label{eq:DecayRate-n-to-llnu}
\end{align}
Here again, $\beta_h \simeq 0.014$ GeV$^3$ is the nuclear matrix
element \cite{Aoki:2017puj} and the half-life is normalized to
the current bound from SNO+ \cite{SNO:2022trz}.
We can apply Eq.~(\ref{eq:DecayRate-n-to-llnu}) to the different existing and
projected half-life limits and obtain estimates for the sensitivity to
the new physics scale $\Lambda$:
\begin{align}
 \Gamma \sim&
 \begin{cases}
  \displaystyle
  \frac{1}{4.9\cdot 10^{26} \text{yr}}
 \left(
 \frac{84\text{TeV}}{\Lambda}
 \right)^{10}
 \text{ Kamiokande (PDG)},
  \\
 %%%%%
  \displaystyle
 \frac{1}{5.0 \cdot 10^{31} \text{yr}}
 \left(
 \frac{270\text{TeV}}{\Lambda}
 \right)^{10}
 \text{ JUNO (future)},
\\
 %%%%%
  \displaystyle
 \frac{1}{4.0\cdot 10^{32} \text{yr}}
 \left(
 \frac{330\text{TeV}}{\Lambda}
 \right)^{10}
 \text{ THEIA (future)}.
 \end{cases}
\label{eq:bound-n-invisible}
\end{align}

\subsection{Variant II: A right-handed neutrino and a ALP \label{subsect:NRA}}

Axion-like particles (ALPs) appear in many extensions of the SM.
While the ALP mass is a free parameter, the couplings of the ALPs with
SM fields are protected by an approximate classical shift symmetry, as
is the case for the classical axion. Many different studies of
properties of and searches for ALPs have been published, see for
example \cite{Brivio:2017ije,Bauer:2021mvw,Bauer:2017ris,
  Biekotter:2025fll}. Here we only briefly mention that ALPs could
also explan the dark matter \cite{Jaeckel:2010ni,Alekhin:2015byh}
and/or be long-lived particles \cite{Alekhin:2015byh,Alimena:2019zri}.

The ALP Lagrangian up to $d=5$ contains the following terms
\cite{Georgi:1986df}
\begin{eqnarray}\label{eq:lag}
  \mathscr{L}_a &=& \frac{1}{2}\partial_{\mu}a\partial^{\mu}a 
             - \frac{1}{2} m_a^2 a^2 
          -\sum_{X} \frac{c_{X \widetilde{X} a}}{\Lambda} 
	  a X^{\mu\nu}\widetilde{X}_{\mu\nu}
	  -\sum_{\psi} \frac{c_{\psi a}}{\Lambda}\partial_{\mu}a
	  ({\overline\psi}\gamma^{\mu}\psi).
\end{eqnarray}
Here $X^{\mu\nu}$ stands for any of the field strength tensors of the
SM, i.e.  $X=B,W,G$, and $\widetilde{X}_{\mu\nu}$ is its dual.  $m_a$ is
the ALP mass, in principle a free parameter. Note, however, that one
usually assumes the ALP to be the pseudo-Goldstones of a spontaneously
broken global symmetry, thus $m_a$ can be naturally small compared to
the scale of symmetry breaking. $\psi$ are the SM fermions and for the
case we are interested in also $\psi=N_R$. It is important to note
that the ALP couples only derivatively to fermions.

There are two $d=8$ $B$-violating operators containing a right-handed
singlet fermion, $N_R$, and an ALP, $a$:
\begin{eqnarray}\label{eq:aNRSMEFT}
  {\cal O}_1^a &=& 
   \frac{c_{\alpha\beta\gamma}^{(1,a)}}{\Lambda^4}
   (\partial_\mu a)
   \left( \overline{N_{R}} \gamma^{\mu} d_{R \alpha} \right)
   \left( \overline{{u_{R \beta}}^{c}} d_{R \gamma} \right),
   \\ 
 \nonumber
   {\cal O}_2^a &=& 
   \frac{c_{\alpha\beta\gamma}^{(2,a)}}{\Lambda^4}
   (\partial_\mu a)
   \left( \overline{N_R} \gamma^{\mu} d_{R \alpha} \right)
   \left( \overline{{Q_{\beta}}^{c}} Q_{\gamma} \right).
\end{eqnarray}
One copy of $N_R$ is enough to form these operators, thus for simplicity
we have not added a generation index to $N_R$. However, two $N_R$ would
still be needed, if one wants to explain neutrino data.

There are two more $d=8$ $B$-violating operators, containing an ALP, that
can be written down with only SM fields:
\begin{eqnarray}
 \label{eq:aSMEFT}
  {\cal O}_3^a &=& 
  \frac{c_{\alpha\beta\gamma\delta}^{(3,a)}}{\Lambda^4}
  (\partial_\mu a)
  \left( \overline{L_{\alpha}} d_{R \beta} \right)
  \left( \overline{{Q_{\gamma}}^{c}} \gamma^{\mu} d_{R \delta} \right),
  \\ 
 \nonumber
  {\cal O}_4^a &=&
  \frac{c_{\alpha\beta\gamma\delta}^{(4,a)}}{\Lambda^4}
  (\partial_\mu a)
  \left( \overline{e_{R \alpha}} \gamma^{\mu} d_{R \beta} \right)
  \left( \overline{{d_{R \gamma}}^{c}} d_{R \delta} \right).
\end{eqnarray}

The operators in Eq.~(\ref{eq:aNRSMEFT}) can generate invisible
neutron decays, without accompanying charged lepton final states,
while ${\cal O}_3^a$ will always also have charged
leptons in the nucleon decays. We have listed $ {\cal O}_4^a$ for
completeness, it does not give a contribution to invisible neutron
decay.

The operators in Eqs.~(\ref{eq:opsNR1}) and (\ref{eq:opsNR2}) violate
$(B+L)=4$. Thus, it is easy to argue that the standard $d=6$ 2-body 
$(B+L)=2$ decays are absent on symmetry grounds for the invisible
neutron decay to three $N_R$'s. However, the situation
is different for the case of an ALP. ALPs do not usually carry
either $B$ or $L$, as can be seen from Eq.~(\ref{eq:lag}). For
an ALP with $(B,L)=(0,0)$, however, ${\cal O}_{1,2}^a$ both have
$(B-L)=2$. One can write down operators at $d=7$ with $(B-L)=2$,
such as, for example, 
${\cal O} \propto (\overline{N_R} \gamma^{\mu} d_{R \alpha})
(\overline{{Q_{\beta}}^{c}} (i D_{\mu}) Q_{\gamma})$. 
Since this type of operators leads to 2-body decays, 
such as $n \to \pi^0 + \esl$
and $p \to \pi^+ + \esl$, one would expect the invisible neutron
decay to be a very sub-dominant decay mode in this case.

A possible way out of this conclusion is based on the following
argument. If we assign $a$ the quantum numbers $(B,L)=(-1,1)$, 
${\cal O}_{1,2}^a$ both conserve $(B+L)$ and $(B-L)$, thus none of the
$(B-L)=2$ operators need to be present in the theory. For this
to be possible, however, the ``standard'' $d=5$ ALP interaction
terms in Eq.~(\ref{eq:lag}) need to be absent. It is possible to
write down UV completions that fulfil these assignments.

For the operators ${\cal O}_{1,2}^a$, the width for the two-body decay
$n \rightarrow N a$ can be estimated to be roughly
\begin{eqnarray}\label{eq:GamNa}
 \Gamma(n \rightarrow N a)
 \simeq
 \frac{\beta_{h}^{2} m_{n}^{3}}{32 \pi \Lambda^{8}}
 \simeq
 \frac{1}{9.0 \cdot 10^{29} \text{yr}}
 \left(
 \frac{9.6 \cdot 10^6 \text{GeV}}{\Lambda}
 \right)^{8}.
\end{eqnarray}
The much larger scale found here, compared to the case $n \rightarrow
3 \bar{N}$, see Eq.~(\ref{eq:DecayRate-n-to-llnu}), makes it unlikely
that the UV completions for these operators can ever be tested in
accelerator experiments.

%%%%%%%%%%%%%%%%%%%%%%%%%%%%%%%%%%%%%%%%%%%%%%%%%%%%%%%%%%%%%%%%%%%%%%
\subsection{Variant III: A right-handed neutrino and a light scalar
  \label{subsect:NRS}}

The third possibility for generating operators for the invisible
neutron decay with BSM fields is to add a right-handed neutrino and a
light scalar to the SM. We discuss this case only briefly, since
we believe it to be less motivated and add it only for completeness of
the discussion.

For a model with a light singlet scalar, $\phi$, and right-handed
neutrinos there are two $d=7$ operators contributing to invisible
neutron decay:
\begin{eqnarray}\label{eq:sNRSMEFT}
  {\cal O}_1^\phi 
   &=& 
   \frac{c_{\alpha\beta\gamma}^{(1,\phi)}}{\Lambda^3}
   \phi
   \left( \overline{{N_{R}}^c} d_{R \alpha} \right)
   \left( \overline{ {u_{R \beta}}^{c}} d_{R \gamma} \right),
   \\ 
 \nonumber
  {\cal O}_2^\phi 
  &=& 
  \frac{c_{\alpha\beta\gamma}^{(2,\phi)}}{\Lambda^3}
  \phi
  \left( \overline{{N_{R}}^c} d_{R \alpha} \right)
  \left( \overline{{Q_{\beta}}^{c}} Q_{\gamma} \right).
\end{eqnarray}
Similar to the ALP case, discussed above, if $\phi$ has $(B,L)=(0,0)$,
the $d=6$ operators that one can construct from
Eq.~(\ref{eq:sNRSMEFT}), eliminating $\phi$, would lead to dominant
2-body decays, rendering invisible neutron decay uninteresting. Asigning
non-zero baryon and lepton number to $\phi$ allows to eliminate this
problem technically.

The width $n \to \bar{N}\phi$ can be estimated to be roughly
\begin{eqnarray}\label{eq:GamNa2}
 \Gamma(n \rightarrow \bar{N} \phi)
 \simeq
 \frac{\beta_{h}^{2} m_{n}}{32 \pi \Lambda^{6}}
 \simeq
 \frac{1}{9.0 \cdot 10^{29} \text{yr}}
% \left(
% \frac{\beta_{h}}{0.014\text{GeV}^{3}}
% \right)^{2}
 \left(
 \frac{2.1 \cdot 10^9 \text{GeV}}{\Lambda}
 \right)^{6}.
\end{eqnarray}

%%%%%%%%%%%%%%%%%%%%%%%%%%%%%%%%%%%%%%%%%%%%%%%%%%%%%%%%%%%%%%%%%%%%%%
\subsection{Variant IV: A right-handed neutrino and a light vector
\label{subsect:NRV}}

Finally, we briefly mention the possibility 
where a neutron decays to a right-handed neutrino and a 
SM-singlet vector boson
$Z'$, which is prompted by the following $d=7$ effective operators:
\begin{eqnarray}\label{eq:ZNRSMEFT}
  {\cal O}_1^{Z'} &=& 
   \frac{c_{\alpha\beta\gamma}^{(1,Z')}}{\Lambda^3}
   Z'_{\mu}
   \left( \overline{N_{R}} \gamma^{\mu} d_{R \alpha} \right)
   \left( \overline{{u_{R \beta}}^{c}} d_{R \gamma} \right),
   \\ 
 \nonumber
   {\cal O}_2^{Z'} &=& 
   \frac{c_{\alpha\beta\gamma}^{(2,Z')}}{\Lambda^3}
   Z'_{\mu}
   \left( \overline{N_R} \gamma^{\mu} d_{R \alpha} \right)
   \left( \overline{{Q_{\beta}}^{c}} Q_{\gamma} \right).
\end{eqnarray}
To avoid the possible $d=6$ nucleon decays and make the $d=7$
invisible neutron decay dominant, we need to assign charges (which
may be the $B$ and $L$ numbers) to $Z'$.
The argument is the same as for the ALP case, cf. Sec.~\ref{subsect:NRA}.

Assuming $Z'$ is a massless gauge boson under a new gauge symmetry,
the decay rate is calculated to be\
\begin{align}
 \Gamma(n \rightarrow N Z') 
 \simeq&
 \frac{\beta_{h}^{2} m_{n}}{16 \pi \Lambda^{6}}
 \simeq
 \frac{1}{9.0 \cdot 10^{29} \text{yr}}
 \left(
 \frac{2.3 \cdot 10^9 \text{GeV}}{\Lambda}
 \right)^{6}.
\end{align}
When $Z'$ aquires a mass through the spontaneous breaking of some new
gauge symmetry, the decay rate changes to:
\begin{align}\label{eq:DecMZ}
 \Gamma \simeq
 \frac{\beta_{h}^{2} m_{n}^{3}}{32 \pi \Lambda^{6} M_{Z'}^{2}}
 \left(
 1 - \frac{M_{Z'}^{2}}{m_{n}^{2}}
 \right)^{2}
 \left(
 1 + \frac{2 M_{Z'}^{2}}{m_{n}^{2}}
 \right).
\end{align}
This equation seems to diverge in the limit of $M_{Z'} \rightarrow 0$.
However, for a gauge boson such divergencies are known to be
unphysical. In a full-fledged model for the massive $Z'$, one would
need to specify also the symmetry breaking sector of the theory. The
full calculation then includes the correct treatment of the would-be
Goldstone boson of the theory and in the limit of $M_{Z'} \ll m_n$
Eq.~(\ref{eq:DecMZ}) will reduce to the same expression as for the
scalar case Eq.~\eqref{eq:GamNa2}, up to a coefficient. However,
constructing such a complete model for the $Z'$ is beyond the scope
of the current paper.

%%%%%%%%%%%%%%%%%%%%%%%%%%%%%%%%%%%%%%%%%%%%%%%%%%%%%%%%%%%%%%%%%%%%%%
\subsection{A caveat \label{subsect:cvt}}

\begin{figure}[t]
 \unitlength=1cm
 \begin{picture}(10,4.5)
 \put(0,0){\includegraphics[width=10cm]{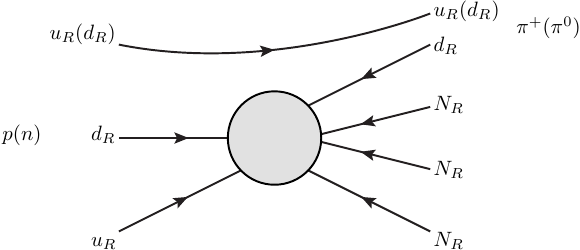}}
 \end{picture}
 \caption{$p(n) \rightarrow \pi^{+}(\pi^{0})+\text{missing}$ induced by
 the $d=9$ $u_{R}d_{R}d_{R}N_{R}N_{R}N_{R}$ operator.}
 \label{Fig:N-to-pi3N}
\end{figure}

There is one important caveat to the above discussion: For any
operator that annihilates three quarks to invisible particles it is
possible to replace an initial state quark for a final state
anti-quark. An example is shown in Fig.~\ref{Fig:N-to-pi3N}.  Thus,
four-body final states with a (neutral or charged) pion will always
accompany three-body invisible decays. - And similarly for the
two-body final states $N_R+a$ and $N_R+S$, three-body final states
with one additional pion will be generated.\footnote{Since one can
replace an initial $d$-quark by a final state $\overline{s}$,
producing a final state kaon, this allows, in principle, also
sensitivity to 2nd generation indices in the operators.}

No search for such three- and four-body final states exist. However,
Super-Kamiokande has a search for $n\to \pi^{0} \bar{\nu}$ and $p \to \pi^{+}
\bar{\nu}$.  The current bounds are \cite{Super-Kamiokande:2013rwg}:
\begin{align}
\Gamma(p \rightarrow \pi^{+}  \bar{\nu}) >& 3.9 \cdot 10^{32} \text{yr},
\\
\Gamma(n \rightarrow \pi^{0}  \bar{\nu}) >& 1.1 \cdot 10^{33} \text{yr}.
\end{align}
These are two-body decays and thus the pion momentum is fixed at roughly 
$p(\pi) \sim 460$ MeV. Super-Kamiokande uses this constraint in setting the
limits. For the 4-body decays, on the other hand, the pion momentum
takes a distribution, which depends in addition on the mass of the
final state particles. Nevertheless, Super-Kamiokande provides sufficient
information such that we can make a rough estimate of the limit for
the 4-body decay $n \to \pi^{0} + 3 \bar{N}$.

The limit on $\Gamma(n \rightarrow \pi^{0} \bar{\nu})$ is set by
Super-Kamiokande \cite{Super-Kamiokande:2013rwg} excluding 19.1 signal
events in the bin $p(\pi^0)=[400,500]$ MeV. From Fig.~3 of
\cite{Super-Kamiokande:2013rwg} we can estimate that the total number
of background events in the range $p(\pi^0)=[0,500]$ MeV sums to about
850 events, and about 80 events in the bin with $p(\pi)=[400,500]$ MeV. 
Thus, we estimate that a limit of $19.1\times
\sqrt{850/80} \simeq 62$ events could be set from the total data.
This would lower the exclusion limit by a factor of roughly $3.3$ and
the limit on $\Gamma(n \rightarrow \pi^{0} + 3 \bar{N}) $ should
roughly be $T_{1/2} \gsim 3.3 \times 10^{32}$ yr. We stress that this
is only a rough estimate and also that the Super-Kamiokande
\cite{Super-Kamiokande:2013rwg} result is based on only $172.8$
kton.yr of data. A re-analysis of all Super-Kamiokande data should be
able to provide a limit, which could be up to a factor two better than
our naive estimate.

%%%%%%%%%%%%%%%%%%%%%%%%%%%%%%%%%%%%%%%%%%%%%%%%%%%%%%%%%%%%%%%%%%%%%%
%%%%%%%%%%%%%%%%%%%%%%%%%%%%%%%%%%%%%%%%%%%%%%%%%%%%%%%%%%%%%%%%%%%%%%

For the four-body decays of Fig.~\ref{Fig:N-to-pi3N}, we can
estimate the partial half-life, using results from
\cite{Heeck:2019kgr}, with appropriate replacements:
\begin{align}
 \Gamma(n \rightarrow \pi^{0} \bar{N}_{i}\bar{N}_{j}\bar{N}_{k})
 \simeq&
 \frac{m_{n}^{7} W_{0}(\pi)^{2}}{737280 \pi^{5} \Lambda^{10}}
% =
% \left(
% \frac{W_{0}(\pi)}{W_{0}(K)}
% \right)^{2}
% \Gamma(n \rightarrow K^{+} \mu^{+} e^{-}e^{-})
% \Bigl|_{\text{massless}}
% \\
% \simeq &
% \left(
% \frac{0.134}{0.23^{2}}
% \right)^{2}
%  \frac{1}{2.60 \cdot 10^{29} \text{yr}}
% \left(
% \frac{100 \text{TeV}}{\Lambda}
% \right)^{10}
% \nonumber
% \\
  \simeq 
\frac{1}{3.3 \cdot 10^{32}\text{yr}}
 \left(
 \frac{245 \text{TeV}}{\Lambda}
 \right)^{10}
\label{eq:DecayRate-n-to-piNNN}
\end{align}
Here the matrix element is taken to be 
%$W_{0}(K) = (0.23\text{GeV})^{2}$ and we used 
$\left|W_{0}(\pi)\right| 
= \left| \langle \pi^{0} | (ud)_{R} d_{R} | n \rangle \right| 
%= \left| \langle \pi^{0} | (ud)_{R} u_{R} | p \rangle \right| 
%= 
%\left| \langle \pi^{0} | (ud)_{L} u_{L} | p \rangle \right| 
= 0.134 \text{GeV}^{2}$~\cite{Aoki:2017puj}. 
And again, we assume the Wilson coefficient is equal to one.

Comparing this estimate with with Eq.~\eqref{eq:bound-n-invisible}, we
can see that the sensitivity of this pion mode to the new physics
scale $\Lambda$ is actually better than the sensitivity of the
invisible mode at SNO+ (180 TeV), and only moderately lower than the
future sensitivity estimated for JUNO (270 TeV). Given these numbers,
one must conclude that searches for $n \to \pi^0 +\esl$ provide
important constraints on searches for invisible neutron decay. 
Therefore, we encourage experimentalists to study the 
$n\rightarrow \pi^{0}+$``multiple missing'' mode 
(independently from  
$n \rightarrow \pi^{0} \bar{\nu}$) and provide the bound.
We stress again, that this is not only true for the case of the $d=9$
operator discussed here, but for all operators generating invisible
neutron decay.

% !TEX root = ../InvNDecay.tex
\section{High-energy completions\label{sect:dec}}

In the previous section we have defined the effective operators for
invisible neutron decay for the four possible different model
variants. In this section we will discuss how these operators can be
generated in the UV, after integrating out heavy particles.  Our aim
here is to find systematically {\em all} possible UV models via the
diagrammatic method. Since, however, the basic procedure is the same
for all decompositions, we will discuss here only one decomposition of
the operator ${\cal O}_1^N \propto u_Rd_Rd_RN_RN_RN_R$ in detail.
Full tables for all other operators and decompositions are given in
the appendix.

%%%%%%%%%%%%%%%%%%%%%%%%%%%%%%%%%%%%%%%%%%%%%%%%%%%%%%%%%%%%%%%%%%%%%%
\begin{figure}[t]
 \unitlength=1cm
 \begin{picture}(6,5.5)
  \put(0,0.5){\includegraphics[width=6cm]{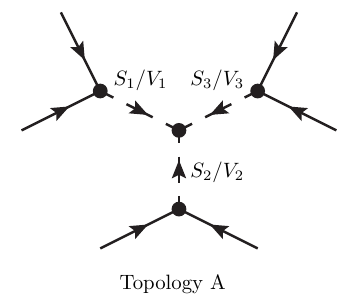}}
 \end{picture}
 \hspace{1cm}
 \begin{picture}(6,5.5)
  \put(0,0.5){\includegraphics[width=6cm]{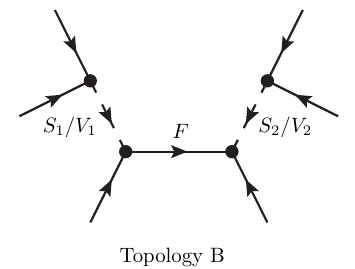}}
 \end{picture}
 \caption{Topologies for tree-level realizations of the $d=9$ neutron
   decay operators, see Eqs.~\eqref{eq:opsNR1} and \eqref{eq:opsNR2}.
   Depending on the chiralities of the outer fermion fields, the
   dashed lines can be either a scalar, $S$, or vector, $V$.  The mediator
   $F$ in Topology B must be introduced as a vector-like fermion, but
   when it is a SM singlet, it can also be a Majorana fermion.  }
\label{Fig:topology-d9}
\end{figure}
%%%%%%%%%%%%%%%%%%%%%%%%%%%%%%%%%%%%%%%%%%%%%%%%%%%%%%%%%%%%%%%%%%%%%%

For 6-fermion operators there are only two possible topologies at
tree-level \cite{Bonnet:2012kh}, using only renormalizable
interactions. They are shown in Fig.~\ref{Fig:topology-d9}. Note that
here the arrows indicate the flow of the particle number. Internal
particles can be either scalars, fermions or vectors. For Topology A
there are exactly three possibilities to distribute the fermions of
${\cal O}_1^N$ to the outer legs of the diagram. We list these in
Tab.~\ref{Tab:decomposition-A}.

\begin{table}[t]
\begin{tabular}{cccccc}
 \hline \hline
 Decomposition & $S_{1}$ & $S_{2}$ & $S_{3}$ & Comment & Eq.
		 \\
 \hline
 $(u_{R} d_{R})(d_{R} N_{R i}) (N_{R j} N_{R k})$
 &$(\overline{\vec{3}}, \vec{1}, +1/3)$
     &$(\vec{3}, \vec{1}, -1/3)$
	 &$(\vec{1},\vec{1},0)$
	     & $S_{2} \neq S_{1}^{\dagger}$ to avoid $d=6$ 
		 & \eqref{eq:L-A-ud-dN-NN}
		 \\
 %%%%%
 $(u_{R} N_{R i}) (d_{R} d_{R}) (N_{R j} N_{R k})$
 & $(\vec{3},\vec{1},+2/3)$
     & $(\overline{\vec{3}},\vec{1},-2/3)$
	 & $(\vec{1},\vec{1},0)$
	     & $d_{R} d_{R} S_{2}^{\dagger} = 0$
		 & 
		 \\
 %%%%%
 $(u_{R} N_{R i}) (d_{R} N_{R j}) (d_{R} N_{R k})$
 & $(\vec{3},\vec{1},+2/3)$
     &$(\vec{3},\vec{1},-1/3)$
	 &$(\vec{3},\vec{1},-1/3)$
	     & $S_{3} \neq S_{2}$ 
		 to avoid $S_{1} S_{2} S_{2} = 0$
		 & \eqref{eq:L-A-uN-dN-dN}
		 \\
 \hline \hline
\end{tabular}
\caption{Decompositions for the operator $u_{R} d_{R} d_{R} N_{R i}
  N_{R j} N_{R k}$ allowed with Topology A.
 The fields in a parenthesis form an interaction with the corresponding
 mediator field, $S_{1,2,3}$. For the explicit form of the interaction
 Lagrangians, cf. Eqs.~\eqref{eq:L-A-ud-dN-NN} and \eqref{eq:L-A-uN-dN-dN}.}
 \label{Tab:decomposition-A}
\end{table}
%%%%%%%%%%%%%%%%%%%%%%%%%%%%%%%%%%%%%%%%%%%%%%%%%%%%%%%%%%%%%%%%%%%%%%

For ${\cal O}_1^N$ and Topology A, all mediator fields are Lorentz
scalars (indicated with symbol $S$). The charges of them under
the SM gauge symmetries, in the order $(SU(3)_{c}, SU(2)_{L},U(1)_{Y})$,
are uniquely determined as given in the table.
The interaction Lagrangians for the different cases can be 
written as:
\begin{align}
\mathscr{L}_{\text{A-1}}
 =&
 y_{ud} \epsilon^{IJK}
 (\overline{{u_{R}}^{c}})_{I \dot{a}}
 (d_{R})_{J}^{\dot{a}}
 (S_{1}^{\dagger})_{K}
 +
 y_{dN}
 (\overline{{d_{R}}^{c}})_{I \dot{a}}
 (N_{R i})^{\dot{a}}
 (S_{2}^{\dagger})^{I}
 +
 y_{NN}
 (\overline{{N_{R j}}^{c}})_{\dot{a}}
 (N_{R k})^{\dot{a}}
 S_{3}^{\dagger}
 \nonumber
 \\
 %%%%%
 &
 +
 \mu
 (S_{1})^{I}
 (S_{2})_{I}
 S_{3},
 \label{eq:L-A-ud-dN-NN}
 \\
 %%%%%
 \mathscr{L}_{\text{A-3}}
 =&
 y_{uN} 
 (\overline{{u_{R}}^{c}})_{I \dot{a}}
 (N_{R i})^{\dot{a}}
 (S_{1}^{\dagger})^{I}
 +
 y_{dNS_2} 
 (\overline{{d_{R}}^{c}})_{I \dot{a}}
 (N_{R j})^{\dot{a}}
 (S_{2}^{\dagger})^{I}
 +
 y_{dNS_3} 
 (\overline{{d_{R}}^{c}})_{I \dot{a}}
 (N_{R k})^{\dot{a}}
 (S_{3}^{\dagger})^{I}
 \nonumber
 \\
 %%%%
 &
 +
 \mu
 \epsilon^{IJK}
 (S_{1})_{I} 
 (S_{2})_{J}
 (S_{3})_{K},
\label{eq:L-A-uN-dN-dN}
\end{align}
where the indices $I,J,K$ in the lower (upper) position are for
$\vec{3}$ ($\overline{\vec{3}}$) in $SU(3)_{c}$, and $\dot{a}$ is for
a right-handed 2-spinor. For case A-2 the interaction
$d_Rd_RS_2^\dagger$ vanishes identically, thus it is not a valid
decomposition for ${\cal O}_1^N$.\footnote{%
However, the interaction $d_Rs_RS_2^\dagger$ does not
vanish.  Since operators are usually given in the flavour basis, while
in the neutron the quarks are in the mass eigenstate basis,
decomposition A-2 can contribute to the invisible neutron decay
rate. The rate, however, will be suppressed by a factor
$\sin^2\theta_C$, where $\theta_C$ is the Cabibbo angle. We will not
discuss this possibility in further details.}  For a more compact
notation, we have suppressed generation indices in these
Lagrangians. The different Yukawas are to be understood as matrices of
either dimensions ($3,3$), ($3,n$) or ($n,n$), where $n$ is the number
of copies of right-handed neutrinos.

A few more comments are in order.  In case A-1, described by the
Lagrangian Eq.~\eqref{eq:L-A-ud-dN-NN}, the diquark, $S_{1}$, has the
same SM charges as the leptoquark, $S_{2}^{\dagger}$.  However, these
two fields have to be distinct states.  If they were identical, one
could construct from $S_{1}$ a $d=6$ effective operator ${\cal O}
\propto u_{R} d_{R} d_{R} N_{R}$. This operator generates the the 2-body
decays $n (p) \rightarrow \pi^{0} (\pi^{+}) \bar{N}$ with a
significantly larger rate than that of the invisible neutron decay.
To make $n \rightarrow 3\bar{N}$ the dominant decay mode, one must
avoid all $d=6$ nucleon decays, i.e. $S_1$ and $S_2$ must be two
different fields. This can be realized by assigning for case A-1 all
$S_i$ definite baryon and lepton numbers. With the asignments $S_1
=(2/3,0)$, $S_2 =(1/3,1)$ and $S_3 =(0,2)$, where the brackets stand
for ($B,L$), all Yukawa couplings conserve baryon and lepton number
trivially. The only $B$ and $L$ violating coupling would then be the
soft parameter $\mu$.
Since $\mu$ is a soft parameter, one can actually argue that it can be
technically small in the sense of 't Hooft's naturalness, since in the limit of
$\mu\to 0$ baryon and lepton numbers are conserved. A small value of
$\mu$ would change the expected mass scale for the $S_i$
drastically. Instead of Eq.~(\ref{eq:DecayRate-n-to-llnu}),
for topology-A one can express the decay width as:
\begin{align}
 \Gamma(n \rightarrow \bar{N}_{i} \bar{N}_{j} \bar{N}_{k})
 \simeq
 \frac{\beta_{h}^{2} m_{n}^{5}\mu^2}{6144 \pi^{3} \Lambda^{12}}
 \simeq
 \frac{1}{9 \cdot 10^{29} \text{yr}}
 \left(
 \frac{1 \text{keV}}{\mu}
 \right)^{2} 
 \left(
 \frac{2.4 \text{TeV}}{\Lambda}
 \right)^{10}.
 \label{eq:DecCA}
\end{align}
Here, it is assumed that $m_{S_1}\simeq m_{S_2}\simeq m_{S_3}\simeq
\Lambda$.  Putting $\mu=\Lambda$, Eq.~(\ref{eq:DecCA}) reduces to
Eq.~(\ref{eq:DecayRate-n-to-llnu}), of course. However, this argument
shows that it might not be completely hopeless to search for
the BSM particles that appear in the decomposition of a $d=9$
operator at the LHC or the future FCC-hh.

The third case, case A-3, is similar to case A-1. However, see
Eq.~\eqref{eq:L-A-uN-dN-dN}, case A-3 contains three leptoquarks, and
two of them, $S_{2}$ and $S_{3}$, interact with $d_{R}$ and
$N_{R}$. They therefore have the same SM charges. However, also in
this case they must be two different fields, otherwise the triple
scalar interaction vanishes. In this case, a simple lepton number
assignment, as discussed for case A-1, is not sufficient to make
$S_{2}$ and $S_{3}$ distinct. This problem can, on the other hand, be
cured by adding a discrete symmetry, for example $\mathbb{Z}_{3}$,
with the charge $\omega\equiv \text{e}^{\text{i}2\pi/3}$. We can
assign $\omega$ to $N_{Rj}$ and $S_{2}$ and $\omega^{2}$ to $N_{R k}$
and $S_{3}$. Under this assignment both, the triple scalar interaction
as well as the Yukawas, are allowed under the discrete symmetry.
However, the $\mu$-term will still break lepton and baryon numbers
softly, thus guaranteeing that the dangerous 2-body nucleon decay
modes are absent.

In the appendix, we list all mediator fields that appear in the
high-energy completions of the $u_{R} d_{R} d_{R} N_{R} N_{R} N_{R}$
operator with Topology B in Fig.~\ref{Fig:topology-d9}.  There we also
give the decompositions of the $QQd_{R} N_{R} N_{R} N_{R}$ operator
Eq.~\eqref{eq:opsNR2} with both the topologies. We note that also for
topology B it is, of course, possible to lower the expected mass
scale. In EFT one usually estimates the scale $\Lambda$ for a Wilson
coefficient $c \simeq 1$. However, for a $d=9$ operator $c \propto
Y^4$, where $Y$ stands symbolically for any Yukawa coupling, see
Fig.~\ref{Fig:topology-d9}. Thus, even moderatly small Yukawa
coouplings, will change the expections given in
Eq.~(\ref{eq:DecayRate-n-to-llnu}) drastically:
\begin{align}
 \Gamma(n \rightarrow \bar{N}_{i} \bar{N}_{j} \bar{N}_{k})
 \simeq
 \frac{\beta_{h}^{2} Y^8 m_{n}^{5}}{6144 \pi^{3} \Lambda^{10}}
 \simeq
 \frac{1}{9 \cdot 10^{29} \text{yr}}
 \left(
 \frac{Y}{0.01}
 \right)^{8} 
 \left(
 \frac{4.5 \text{TeV}}{\Lambda}
 \right)^{10}.
 \label{eq:DecCB}
\end{align}
In addition, the appendix contains tables of the decompositions for
the effective operators in the other variants, which were presented in
Secs.~\ref{subsect:NRA}-\ref{subsect:NRV}. Similar to cases A-1
and A-3 discussed here for ${\cal O}_1^N$, in all cases one has to
make sure, that the dangerous 2-body nucleon decays are absent by the use of
some symmetry, otherwise the rate for the invisible neutron decay will
be negligible.

%%%%%%%%%%%%%%%%%%%%%%%%%%%%%%%%%%%%%%%%%%%%%%%%%%%%%%%%%%%%%%%%%%%%%%
\begin{table}[t]
 \begin{tabular}{ccccccc}
  \hline \hline
  & \multicolumn{2}{c}{$n\rightarrow 3\bar{N}$}
      & \multicolumn{2}{c}{$n\rightarrow N a (NZ')$}
  & \multicolumn{2}{c}{$n\rightarrow \bar{N} \phi$}
  \\
  Mediator 
  & $uudNNN$ \hspace{0.5cm}
      & $QQdNNN$ \hspace{0.5cm}
	  & \begin{minipage}{1.5cm}
	     $\partial a udd\bar{N}$
	     \\
	     \hspace*{-0.1cm}$(Z' udd\bar{N})$
	    \end{minipage}
  \hspace{0.5cm}
	      & 
  \begin{minipage}{1.5cm}
   $\partial a QQd\bar{N}$
   \\
   \hspace*{-0.2cm}$(Z' QQd\bar{N})$
  \end{minipage}
  \hspace{0.5cm}
  & $uudN\phi$
      \hspace{0.5cm}
  & $QQdN\phi$
  \\
  %%%%%
  \hline
  $S(\vec{1},\vec{1},0)$
  & $\checkmark$
      & $\checkmark$
	  \\
  %%%%%
  $S(\vec{3},\vec{1},-1/3)$
  & $\checkmark$
       & $\checkmark$
	  & $\checkmark$
	      & $\checkmark$
		  & $\checkmark$
		      & $\checkmark$
      \\
  %%%%%
  $S(\vec{3},\vec{1},+2/3)$
  & $\checkmark$
      & & $\checkmark$
	      & & $\checkmark$ 
  &
   \\   
  %%%%%
  $S(\vec{3},\vec{2},+1/6)$
  & & & & $\checkmark$ & & 
  \\
  %%%%%
  \hline
  $V(\vec{3},\vec{1},-1/3)$
  & & & $\checkmark$ & $\checkmark$ & &
  \\
  %%%%%
 $V(\vec{3},\vec{1},+2/3)$
  & & & $\checkmark$
  \\
  %%%%%
 $V(\vec{3},\vec{2},+1/6)$
  & & $\checkmark$ & & $\checkmark$ & & $\checkmark$ 
  \\
  %%%%%
  \hline
  $F(\vec{1},\vec{1},0)$
  & $\checkmark$
      & $\checkmark$
	  & $\checkmark$
	      & $\checkmark$
  & $\checkmark$
  & $\checkmark$
  \\
  %%%%%
  $F(\vec{3},\vec{1},-1/3)$
  & $\checkmark$
      & $\checkmark$
	  & $\checkmark$
	      & $\checkmark$
  & $\checkmark$
  & $\checkmark$
  \\
  %%%%%
  $F(\vec{3},\vec{1},+2/3)$
  & $\checkmark$ &
	  & $\checkmark$ &
  & $\checkmark$ &
  \\
  %%%%%
  $F(\vec{3},\vec{2},+1/6)$
  & & $\checkmark$
	  & & $\checkmark$
  & & $\checkmark$
  \\
  \hline \hline
\end{tabular}
\caption{List of the mediators that appear in the decompositions of the
 effective operators relevant for the invisible neutron decay.}
\label{Tab:mediator-list}
\end{table}
%%%%%%%%%%%%%%%%%%%%%%%%%%%%%%%%%%%%%%%%%%%%%%%%%%%%%%%%%%%%%%%%%%%%%%

Table \ref{Tab:mediator-list} lists all mediator fields that appear
in the decompositions of the $d=(7-9)$ operators. There is a total of
11 fields, 4 scalars, 4 fermions and 3 vectors. Some fields appear in
all operators and in all 4 model variants ($S_{3,1,-1/3}$, $F_{1,1,0}$
and $F_{3,1,-1/3}$), while some others appear only for a particular
model variant ($S_{1,1,0}$, $V_{3,1,-1/3}$) or even only one
particular operator ($S_{3,2,1/6}$ and $V_{3,1,2/3}$). Note that the
fields appearing in the decompositions of the effective operator with
a $Z'$ are the same as the ones for the operators with an ALP, the
corresponding column applies to both fields.

% !TEX root = ../InvNDecay.tex
\section{LHC phenomenology\label{sect:pheno}}

In this section, we will briefly discuss a variety of LHC searches,
that can be re-interpreted as limits for fields listed in 
Tab.~\ref{Tab:mediator-list}. Naive expectations for the scale, $\Lambda$,
of the $d=(7-9)$ operators, discussed in secton \ref{sect:ops}, make
it look quite unlikely that the LHC can find any positive signals for
these states. However, as discussed around Eqs.~(\ref{eq:DecCA}) and
(\ref{eq:DecCB}), the estimates derived using EFT in section
\ref{sect:ops} might vastly overestimate the masses of the BSM fields in
a full-fledged UV model. We therefore felt motivated to at least
briefly discuss the current experimental status.

The decomposition of the $d=9$ operators for the $N_R$SMEFT model,
discussed in section \ref{sect:dec}, generates a set of eight
different BSM fields: Four vector-like fermions ($F_{1,1,0}$,
$F_{3,1,-1/3}$, $F_{3,1,2/3}$ and $F_{3,2,1/6}$), three types of
scalars ($S_{1,1,0}$, $S_{3,1,-1/3}$ and $S_{3,1,2/3}$) and one vector
($V_{3,2,1/6}$). Three more fields appear only in the ALP/$Z'$
variants ($S_{3,2,1/6}$, $V_{3,1,-1/3}$ and $V_{3,1,2/3}$). We mention
these only for the sake of completeness here, since a LHC discovery
for the fields from the $d=7,8$ operator seems (even) less likely than
those appearing in models for the $d=9$ operators.

First, we mention that for the singlet fields, there is no production
mode at the LHC other than decays of the coloured BSM
states. Moreover, $F_{1,1,0}$ and $S_{1,1,0}$ will decay to $N_R$'s
and thus be invisible, if they are lighter than the coloured
states. Thus, there are no constraints on these two fields from LHC
searches. 

\begin{figure}[t!]
    \centering
    \includegraphics[scale=0.7]{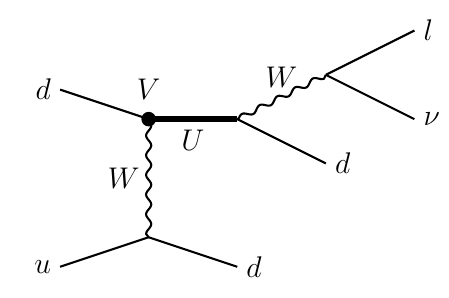}\hskip5mm
    \includegraphics[scale=0.7]{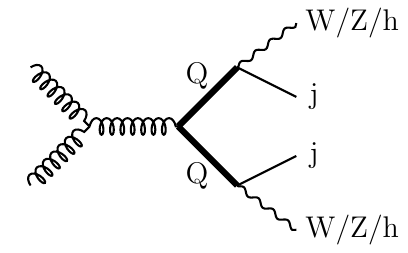}
    \caption{Example Feynman diagrams for single and pair 
      production of VLQs and their decays.}
    \label{fig:directprod}
\end{figure}

Let us then concentrate on the coloured fermions. Here, we follow
largely the discussion in \cite{Cepedello:2024qmq}. All three coloured
fermions can have interactions with SM quarks and the Higgs:
\begin{eqnarray}\label{eq:lagY}
{\cal L}_{\rm Y} &=& Y_{U_R} \overline{Q} F_{3,1,2/3} H^{\dagger}
          +  Y_{D_R} \, \overline{Q} F_{3,1,-1/3} H
          +  Y_{Q_u} \, \overline{F_{3,2,1/6}} u_R H^{\dagger}
          +  Y_{Q_d} \, \overline{F_{3,2,1/6}} d_R H  + {\rm h.c.} 
\end{eqnarray}          
In Eq. (\ref{eq:lagY}) we have suppressed generation indices.  The
fermions $F_{3,1,-1/3}$, $F_{3,1,2/3}$ and $F_{3,2,1/6}$ must also
have vector-like mass terms. After electro-weak symmetry breaking,
Eq.~(\ref{eq:lagY}) leads to mixing between heavy and SM quarks. This
mixing will dominate their decay widths, unless the corresponding
Yukawa couplings are tiny.  If, on the other hand, some Yukawa
couplings are large, this mixing may dominate also VLQ production at
the LHC, see Fig.~\ref{fig:directprod}, left. In addition, VLQs can
always be pair-produced, see Fig.~\ref{fig:directprod} to the right,
independent of the size of the mixing. Note that in this figure we 
show the VLQs decaying to ``j'' to indicate any final state quark,
i.e. including also third generation.

Both ATLAS and CMS have searched for such vector-like quarks in a
number of publications. For example, ATLAS has searched for
pair-produced VLQs decaying to 1$^{\rm st}$ generation quarks in
\cite{ATLAS:2024zlo}. Limits are given as function of the VLQ mass,
for different branching ratios Br($Q \to W + j$) versus Br($Q \to H +
j$), assuming Br($Q \to W + j$)+Br($Q \to H + j$)+Br($Q \to Z +
j$)$=1$ \cite{ATLAS:2024zlo}. Derived limits are in the range
$(900-1500)$ GeV for Br($Q \to W + j$) in the range (10-100) \%. 
In \cite{ATLAS:2024gyc} ATLAS presented the results of a search
for pair-produced VLQs decaying to third generation quarks.
Limits range from $(1-1.7)$ TeV depending mostly on the
branching ratio Br($Q \to W + t/b$). These limits are slightly
stronger than the 1$^{\rm st}$ generation quark limits, mostly
due to lower backgrounds.

There are also searches for singly produced VLQs.  However, in the
search \cite{ATLAS:2022ozf} ATLAS assumes that the VLQ decays to the
final state $t+h$ or $W + b$, where the top/bottom is tagged in order
to reduce backgrounds. Similarly CMS \cite{CMS:2024qdd} searched for
single VLQ production in the final state $t+h$ and $t+Z$. For
$\kappa_T=1$, where $\kappa_T$ measures the strength of the
interaction of the VLQ with electro-weak gauge bosons relative to
standard EW couplings, a lower limit on VLQs decaying to top quarks of
around $m_{\rm VLQ} \ge 1.9$ TeV is found in \cite{ATLAS:2022ozf}.
There are no searches yet published for singly produced VLQs decaying
to second or 1$^{\rm st}$ generation quarks. However, due to much larger
backgrounds in such searches, one can expects limits would be considerably
weaker. 

However, we need to stress that all the VLQ searches discussed up to
now depend on the presence of mixing.  The neutron decay diagrams,
that we study here, however, do not require any of the Yukawa coupling
in Eq.~(\ref{eq:lagY}) to be non-zero. Instead, couplings of the form
$F$-$q$-$S$, where $F$, $q$ and $S$ stand for a VLQ, a SM quark and one of
the BSM scalars must be present. If the VLQs are pair produced at the
LHC they can then decay to a final states containing $q+N_R+N_R$,
i.e. jet plus missing energy.  This signal is equivalent to the one
used in the standard squark search at the LHC. For example, ATLAS
\cite{ATLAS:2020syg} has searched for pair produced squarks with
${\cal L}=139/$fb. ATLAS derives limits for one generation of squarks
(roughly $m_{\tilde q}\ge 1.2$ TeV) and 8 degenerate squarks (roughly
$m_{\tilde q}\ge 1.8$ TeV) for light neutralinos. Since VLQs (being
fermions) have cross sections at least $\sim$ 4 times larger than a
colour triplet scalar of the same mass, this SUSY search
\cite{ATLAS:2020syg} could be reinterpreted to yield a lower limit on
the VLQ mass. We estimate this limit to be roughly $m_{\rm VLQ} \ge
1.5$ TeV, for any of the three coloured VLQs under consideration.

The coloured scalar states appear either as leptoquarks or as
diquarks in the different decompositions, discussed in the
previous sections. The leptoquark couplings are:
\begin{eqnarray}\label{eq:LQ}
  {\cal L}^{\rm LQ} = & & Y_{Nu}\ 
   \overline{{u_R}^c}N_R S^{\dagger}_{3,1,2/3}
   + Y_{de}\ \overline{{d_R}^{c}} e_R S_{3,1,2/3} \\ \nonumber
   & +& Y_{Nd}\ \overline{{d_R}^c}N_R S^{\dagger}_{3,1,-1/3}
                  + Y_{eu}\ \overline{{u_R}^c}e_R S^{\dagger}_{3,1,-1/3}
                  + Y_{LQ}\ \overline{Q^c}L S^{\dagger}_{3,1,-1/3}
                  + {\rm h.c.},
\end{eqnarray}  
while the diquark couplings can be written as:
\begin{equation}\label{eq:DQ}
  {\cal L}^{\rm DQ} = Y_{dd}\ \overline{{d_R}^c}d_R S_{3,1,2/3}
                  + Y_{QQ}\ \overline{Q^c}Q S_{3,1,-1/3}
                  + Y_{ud}\ \overline{{u_R}^c}d_R S_{3,1,-1/3}
                  + {\rm h.c.}
\end{equation}
Again, we have suppressed generation indices. We stress, however, that
the flavour diagonal entries in $Y_{dd}$ are zero, see section
\ref{sect:dec}.  Importantly, while the LQs and DQs appear with the
same SM quantum numbers, they {\em must} be different states,
otherwise one can generate $d=6$ proton decay tree-level diagrams,
which would lead to lower limits on their masses near the GUT scale
and render invisible neutron decay completely negligible, as discussed
in section \ref{sect:dec}. Note that only ($Y_{Nu}$, $Y_{Nd}$ and
$Y_{ud}$) of the eight couplings given in Eqs.~(\ref{eq:LQ}) and
(\ref{eq:DQ}) appear in the decompositions for the neutron decay. This
is important, since none of these couplings will generate final states
with charged leptons.

Leptoquarks, both pair and singly produced, have been searched for at
the LHC in a number of different final states. The couplings $Y_{eu}$
and $Y_{LQ}$ do not appear in the decompositions, but will lead to
final states with charged leptons, that are more easily constrained at
the LHC. For example, \cite{ATLAS:2024huc} searched for pair produced
LQs decaying to 3$^{\rm rd}$ generation quarks. Lower limits on LQ
masses in the range $(1.2-1.7)$ TeV are derived, depending on LQ decay
branching ratios.  ATLAS also searched for LQs decaying to light
quarks \cite{ATLAS:2020dsk}.  For a branching ratio of Br(LQ$\to l +
q$)$=1$ limits of $m_{LQ} \ge 1.8$ TeV and $m_{LQ} \ge 1.7$ TeV are
excluded in the electron and muon channels. These limits weaken
to roughly $(700-900)$ GeV for Br(LQ$\to l +q$)$=0.1$.

Single or resonant production of LQs requires large Yukawa couplings
to give sizeable cross sections. ATLAS searched for resonant LQs
\cite{ATLAS:2025upm} via a lepton-jet signature, with either one or
two leptons in the final state.  For $Y_{de} =1.0$ LQs with masses
below $m_{LQ}=3.4$ TeV are excluded by this search.  CMS searched for
single LQs in t-channel diagrams \cite{CMS:2025iix}, leading to
di-lepton final states.  Both scalar and vector LQs were
considered. For scalar [vector] LQs with masses between $(1-5)$ TeV
Yukawa couplings larger than $(0.3–1.0)$ [$(0.1-1.4)$] have been
excluded.

Again, the searches just discussed require charged lepton final states
and the corresponding couplings are not required by the decompositions
for invisible neutron decay.  LQs that decay only to quarks and
invisible final states are less constrained. For example, the SUSY
search by ATLAS \cite{ATLAS:2020syg} will give a limit of roughly
$m_{LQ} \gsim 1.2$ TeV for pair produced LQ.  A CMS paper
\cite{CMS:2019ybf} gives limits for pair produced LQs decaying to
neutrinos. Limits are $m_{LQ} \ge 1140$ GeV for scalar LQs and $m_{LQ}
\ge (1560-1980)$ GeV for vector LQs, depending on the strength of the
LQ gluon coupling. The more stringent limits for vector LQs simply
reflect the larger cross sections for vectors compared to
scalars. These limits apply to the state $V_{3,2,1/6}$ that appears in
the decomposition of ${\cal O}_2^N$.

Finally, diquarks are stringently constrained from dijet searches,
since the production is s-channel enhanced.  For example, CMS
\cite{CMS:2019gwf} gives a lower limit on the mass of a colour triplet
scalar of $m_{DQ} \simeq 7.5$ TeV assuming a Yukawa coupling to two
quarks of electromagnetic strength, $e$. The limits derived by CMS are
strictly speaking valid only for a scalar diquark coupling to both,
up-type and down-type quarks with the same strength. Thus, they apply
only to $S_{3,1,-1/3}$.

The searches discussed above are based on luminosities up to 140/fb.
Moderate improvements can be expected from the high-luminosity LHC.
However, the future FCC-hh  \cite{FCC:2018vvp} would considerably
improve sensitivities. For example, \cite{FCC:2018byv} quotes
sensitivities up to $m_{LQ} \sim 8$ ($15$) TeV for pair (singly)
produced leptoquarks, decaying to charged leptons. Many other search
channels, discussed above, should improve by similar factors.

We will close this section with a short discussion of the $d=12$
operator given in Eq.~(\ref{eq:W12}). In the beginning of this section
we have argued that a Wilson coefficient of $c_W=1$ is often
unrealistic in UV models, thus EFT tends to overestimate the
masses of the BSM states. This is certainly true also for a $d=12$
operator.\footnote{Note that in the decomposition of
Eq.~(\ref{eq:W12}) up to seven different Yukawa couplings appear.}  We
have used an automated code for operator decomposition, based on the
diagrammatic method and described in \cite{Cepedello:2022pyx,
  Cepedello:2023yao}, to decompose Eq.~(\ref{eq:W12}) at tree-level.
At such a large dimension, there is a proliferation of models. We
count 39 topologies, 150 different diagrams and 12713 model
variants.

Despite this large number of possible model variations all model
variants can be described with just 38 different fields (scalars and
fermions, no vectors). 26 of these are triplets of colour, either
diquarks, leptoquarks or heavy vector-like quarks.  All scalars and
fermions discussed above appear in this list too and the constraints
we have discussed above apply also to this case. The remaining
coloured states, not covered in the above discussion will have bounds,
which are at least as strong, since they are all either larger $SU(2)$
multiplets or states with larger hyper-charge. There is, however, one
important difference between the UV states of the $d=12$ operator and
those of the $d=(7-9)$ operators. Whereas for the $d=9$ operators,
strictly speaking, only right-handed neutrino final states are
required by the couplings appearing in the decomposition, in the case
of the $d=12$ operator similar decay rates of the leptoquarks to final
states with missing energy and final states with charged leptons are
expected. As discussed above, final states with charged leptons are a
better signal than missing energy at the LHC and thus give generally
more stringent limits. In other words, limits on states from the
decompositions of the $d=12$ operator from LHC will in general be more
stringent than those for the $d=9$ case, that motivated the discussion 
given in this section.

% !TEX root = ../InvNDecay.tex
\section{Summary \label{sect:cncl}}

In this paper we have studied invisible neutron decay. Limits on
nucleon decay modes with three charged leptons
\cite{Super-Kamiokande:2020tor} are much stronger than those of
invisible neutron decay \cite{ParticleDataGroup:2024cfk}. Invisible
neutron decay with observable rates can therefore be generated only by
operators that do not also generate charged lepton final states.  In
SMEFT one has to go up to $d=12$ to find such an operator.  Current
limits on invisible neutron decay correspond then to scales of order
$\Lambda \simeq 13 $ TeV. An observation of invisible neutron
decay in the next round of experiments might then be accompanied by
new physics at the future FCC-hh \cite{FCC:2018byv}.

However, this conclusion changes, if one adds new light degrees of
freedom to the SM particle content. We have discussed four different
extensions of the SM for which invisible neutron decay with observable
rates can be induced by operators of dimension $d=(7-9)$. We have
estimated the half-lives in each case and for each model we also give
all possible UV completions at tree-level. Current limits on the
scales of these operators range from $\Lambda \simeq 180$ TeV for
the variant model with only right-handed neutrinos ($d=9$ operator)
to $\Lambda \simeq 2\cdot 10^9$ GeV ($d=7$ operator).

A discovery of invisible neutron decay, therefore may point towards to
existence of new, light BSM states. In particular, we stress that in
all possible models a neutral fermion state with the quantum numbers
of a right-handed neutrino must exist. Thus, these models will also be
able to explain neutrino masses.

We have also discussed constraints on the allowed rate for invisible
neutron decay, that could be placed by a search for $n \to \pi^0
+\esl$. Hyper-Kamiokande \cite{Hyper-Kamiokande:2018ofw} should be
able to improve on the limit of $n \to \pi^0 +\esl$ from
Super-Kamiokande \cite{Super-Kamiokande:2013rwg} by a considerable
factor. We note that the Super-Kamiokande analysis
\cite{Super-Kamiokande:2013rwg} gives only a limit on a two-body final
state, but the same type of search can also be used to constrain
decays with more than one invisible particle in the final state.
Non-observation of $n \to \pi^0 + 3 \bar{N}$ at Hyper-Kamiokande might
rule out the possibility of observing invisible neutron decay in
the foreseeable future.

However, to end this discussion on a more positive note, we mention
that the sensitivity of Hyper-Kamiokande and the future JUNO and (the
proposed) THEIA experiment to the scales of the invisible neutron
decay operators will be rather similar. Observation of invisible
neutron decay in JUNO should lead to an excess in the Hyper-Kamiokande
search for $n \to \pi^0 +\esl$ and vice versa.  The question whether
invisible neutron decay or the decay $n \to \pi^0 +\esl$ would be {\em
  discovered first}, however, is not so easy to answer.  While the
future Hyper-Kamiokande experiment \cite{Hyper-Kamiokande:2018ofw} is
much larger than Super-Kamiokande, and thus allows to probe much
larger half-lives, already the Super-Kamiokande limit
\cite{Super-Kamiokande:2013rwg} is background dominated. To claim a
discovery in this case requires both hundreds of signal events and a
detailed understanding of backgrounds.  JUNO's search for the
invisible neutron decay \cite{JUNO:2024pur}, on the other hand, is
based on a characteristic triple coincidence arising from the
invisible decays of s-shell neutrons in $^{12}$C, which leaves very
few background events, making a discovery with just a handful of
events possible.

Finally, one can speculate that adding the two gammas from the $\pi^0$
decay to the triple coincidence for the invisible neutron decay would
eliminate all remaining background in JUNO's search. We do not know,
however, whether this advantage is (over-)compensated by the loss of
signal extraction efficiency or not. Only the experimental
collaboration can perform an analysis sufficiently sophisticated to
decide whether $n \to \pi^0 +\esl$ or the invisible neutron decay
itself would be more sensitive to the different models (and the
corresponding operators) that we discussed in the present paper.

%%%%%%%%%%%%%%%%%%%%%%%%%%%%%%%%%%%%%%%%%%%%%%%%%%%%%%%%%%%%%%%%%%%%%%
\bigskip
\centerline{\bf Acknowledgements}

\medskip

J.C.H and T.O acknowledge support from ANID – Millennium Science
Initiative Program ICN2019 044.  
The research of J.C.H is supported by
ANID Chile through FONDECYT regular grant N${}^{\underline{\text{o}}}$
1241685.
The research of T.O. is supported by
ANID Chile through FONDECYT regular grant N${}^{\underline{\text{o}}}$
1250343. 
M.H. acknowledges support by Spanish grants
PID2023-147306NB-I00 and CEX2023-001292-S
(MCIU/AEI/10.13039/501100011033), as well as CIPROM/2021/054
(Generalitat Valenciana).

%%%%%%%%%%%%%%%%%%%%%%%%%%%%%%%%%%%%%%%%%%%%%%%%%%%%%%%%%%%%%%%%%%%%%%

%%%%%%%%%%%%%%%%%%%%%%%%%%%%%%%%%%%%%%%%%%%%%%%%%%%%%%%%%%%%%%%%%%%%%%
\appendix
\bigskip
\section{List of the high-energy completions}
\label{app}

\begin{figure}[t]
 \unitlength=1cm
 \begin{picture}(5,4.5)
  \put(0,0){\includegraphics[width=5cm]{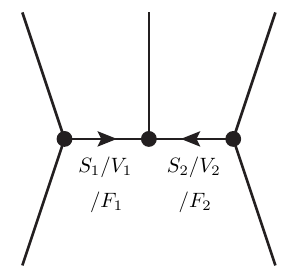}}
   \end{picture}
 \caption{Topology for the operators with four fermions and 1 scalar,
 such as $u_{R} d_{R} d_{R} N_{R} \phi$
 and $u_{R} d_{R} d_{R} \overline{N_{R}} a$,
 i.e.,
 One of the outer legs should be a scalar, $\phi$ or $a$.
 The directions of arrows on the outer legs depends on the decomposition.
 Although the mediators are given with solid lines in this topology, 
 they can be scalars $S$, vectors $V$ or fermions $F$, depending on 
 the distribution of outer fields.}
 \label{Fig:d7-uddNp}
\end{figure}

Here we list the decompositions of the effective operators and the
charges under the SM gauge symmetries of the necessary mediator fields in
each decomposition.
\begin{itemize}
 \item Variant I with light right-handed neutrinos $N_{R}$
       ---
       Tabs.~\ref{Tab:decomposition-A},
       \ref{Tab:decomposition-B},
       \ref{Tab:QQdNNN-decomposition-A},
       and 
       \ref{Tab:QQdNNN-decomposition-B}.
       Fig.~\ref{Fig:topology-d9} for the topologies.
       
 \item Variant II with an $N_{R}$ and an axion-like particle $a$
       ---
       Tabs.~\ref{Tab:uddNa-decomposition}
       and \ref{Tab:QQdNa-decomposition}.
       Fig.~\ref{Fig:d7-uddNp} for the topology.

 \item Variant III with an $N_{R}$ and a scalar field $\phi$
       ---
       Tabs.~\ref{Tab:uddNp-decomposition} and \ref{Tab:QQdNp-decomposition}.
       Fig.~\ref{Fig:d7-uddNp} for the topology.

 \item Variant IV with an $N_{R}$ and a vector boson $Z'$
       ---
       The mediators appear in this variant are the same as 
       those listed in Tabs.~\ref{Tab:uddNa-decomposition}
       and \ref{Tab:QQdNa-decomposition} for Variant II.
\end{itemize}

The ``Decomposition'' column shows 
how the outer fields are distributed to the vertices in the
corresponding topology.
The fields in a parenthesis are in the same vertex.
Let us take
the first line of Tab.~\ref{Tab:decomposition-B}
\begin{align}
 (u_{R} d_{R})(N_{R i})(N_{R j})(d_{R} N_{R k})
\end{align} 
as an example.
``$(u_{R} d_{R})$'' means that $u_{R}$ and $d_{R}$ correspond to 
the two outer fields in the leftmost vertex in Topology B given 
in Fig.~\ref{Fig:topology-d9}.
The mediator field between the $(u_{R} d_{R})$ vertex and the 
second-left vertex with $N_{R i}$ 
is $S_{1}(\overline{\vec{3}},\vec{1},+1/3)$.
The direction of the $S_{1}$ particle number is indicated with arrow in 
Fig.~\ref{Fig:topology-d9}.
The two of the middle vertices with $N_{Ri}$ and $N_{Rj}$
are mediated by the fermion $F(\overline{\vec{3}},\vec{1},+1/3)$.
The outer fields in the rightmost vertex are $d_{R}$ and $N_{Rk}$,
and the mediator between the $(N_{Rj})$ vertex and the $(d_{R} N_{Rk})$
vertex
is $S_{2}(\vec{3},\vec{1},-1/3)$.
In short, the high-energy completion of this decomposition is 
determined as
\begin{align}
 \mathscr{L}_{\text{Tab.\ref{Tab:decomposition-B}-1}}
 =&
 y_{ud}
 \epsilon^{IJK}
 (\overline{{u_{R}}^{c}})_{I \dot{a}}
 (d_{R})_{J}^{\dot{a}}
 (S_{1}^{\dagger})_{K}
 +
 y_{dN}
 (\overline{{d_{R}}^{c}})_{I \dot{a}}
 (N_{R})^{\dot{a}}
 (S_{2}^{\dagger})^{I}
 \nonumber
 \\
 %%%%%
 &
 +
 y_{\overline{F}NS_{1}}
 (\overline{F_{L}})_{I \dot{a}}
 (N_{R})^{\dot{a}}
 (S_{1})^{I}
 +
 y_{NFS_{2}}
 (\overline{{N_{R}}^{c}})_{\dot{a}}
 (F_{R})^{I \dot{a}}
 (S_{2})_{I}.
\end{align}
The charges of $S_{2}$ and $S_{1}^{\dagger}$ are the same.
However,
if they are an identical field,
it mediates the $d=6$ effective operator
\begin{align}
 \mathscr{L}_{d=6}
 =
 \frac{
 y_{ud} 
 y_{dN}
 }{M_{S_{1}}^{2}}
 \epsilon^{IJK}
 (\overline{{u_{R}}^{c}})_{I \dot{a}}
 (d_{R})_{J}^{\dot{a}} 
 (\overline{{d_{R}}^{c}})_{K \dot{b}}
 (N_{R})^{\dot{b}}
\end{align}
which induces $p \rightarrow \pi^{+} \bar{N}$ with 
larger decay rate than $d=9$ induced invisible neutron decay,
i.e.,
the strong bound from $p \rightarrow \pi^{+} \bar{N}$ 
constrains the invisible neutron decay to an 
unreachable level.
In short, if one wants to have this decomposition for the invisible
neutron decay, the mediator $S_{2}$ must be an independent 
field from $S_{1}^{\dagger}$, although the SM charges are the same.
This is mentioned in the ``Comment'' column
as ``$S_{2} \neq S_{1}^{\dagger}$ to avoid $d=6$''. 

The decompositions with the comment ``$d_{R}d_{R}S^{\dagger}=0$''
contain the interaction 
\begin{align}
 \mathscr{L}
 =
 y \epsilon^{IJK} (\overline{{d_{R}}^{c}})_{I \dot{a}}
 (d_{R})_{J}^{\dot{a}}
 (S^{\dagger})_{K}
\label{eq:L-ddS-vanish}
\end{align}
which vanishes because of the colour antisymmetric nature of the two $d_{R}$s
forming a scalar.
The decomposition $(u_{R} N_{Ri})(d_{R})(d_{R})(N_{Rj} N_{Rk})$
does not contain the interaction Eq.~\eqref{eq:L-ddS-vanish},
but the effective operator induced from this decomposition 
results in the same structure $\epsilon^{IJK}(\overline{{d_{R}}^{c}})_{I
\dot{a}} (d_{R})_{J}^{\dot{a}}$ that vanishes.

For all fermion mediators that appear in the decompositions, we assume a
vector-like fermions (left and right 2-spinors with the same charges
form a 4-componenet Dirac fermion).
However, the mediator of the SM singlet field can be a Majorana fermion 
(only 2 components out of 4 are independent).
The decompositions with the SM singlet fermion mediator are indicated 
with the comment ``$F$ can be Majorana''.

In the last decomposition in Tab.~\ref{Tab:QQdNNN-decomposition-A},
the two vector mediators have the same SM charges.
Since they are antisymmetric under both $SU(3)_{c}$ and $SU(2)_{L}$,
the interaction $V_{1} V_{2} S_{3}$ 
\begin{align}
 \mathscr{L}_{\text{Tab.\ref{Tab:QQdNNN-decomposition-A}-3}}
 =
 \epsilon^{IJK}
 (V_{1})_{I i \rho}
 (\text{i}\tau^{2})^{ij}
 (V_{2})_{J j}^{\rho}
 (S_{3})_{K}
 +
 \cdots
\end{align}
does not vanish even if they are identical fields.
This is different from the last decomposition in
Tab.~\ref{Tab:decomposition-A}, where $S_{3} \neq S_{2}$ is required so
that the $S_{1} S_{2} S_{3}$ interaction does not vanish.
The comment ``$V_{1} = V_{2}$'' indicates this fact.

%%%%%%%%%%%%%%%%%%%%%%%%%%%%%%%%%%%%%%%%%%%%%%%%%%%%%%%%%%%%%%%%%%%%%%
\begin{table}[t]
\begin{tabular}{ccccc}
 \hline \hline
 Decomposition & $S_{1}$ & $F$ & $S_{2}$ & Comment %& Eq.
 \\
 %%%%%
 \hline
 $(u_{R} d_{R})(N_{R i})(N_{R j})(d_{R} N_{R k})$
 & $(\overline{\vec{3}}, \vec{1}, +1/3)$
     & $(\overline{\vec{3}},\vec{1},+1/3)$
	 & $(\vec{3},\vec{1},-1/3)$
	     & $S_{2} \neq S_{1}^{\dagger}$ to avoid $d=6$
%		 & \eqref{eq:L-B-ud-N-N-dN}
		 \\
 %%%%%
 $(u_{R} d_{R})(d_{R})(N_{R i})(N_{R j} N_{R k})$
 & $(\overline{\vec{3}}, \vec{1}, +1/3)$
     & $(\vec{1},\vec{1},0)$
	 & $(\vec{1},\vec{1},0)$
	     & $F$ can be Majorana 
%		 & \eqref{eq:L-B-ud-d-N-NN}
%		     \eqref{eq:L-B-ud-d-N-NN-Majo}
		 \\
 %%%%%
 $(u_{R} d_{R})(N_{R i})(d_{R})(N_{R j} N_{R k})$
 & $(\overline{\vec{3}}, \vec{1}, +1/3)$
     & $(\overline{\vec{3}},\vec{1},+1/3)$
	 & $(\vec{1},\vec{1},0)$
	     &
%		 & \eqref{eq:L-B-ud-N-d-NN}
	     \\
 %%%%
 $(d_{R} N_{R i}) (u_{R})(d_{R}) (N_{R j} N_{R k})$
 & $(\vec{3},\vec{1},-1/3)$
     & $(\overline{\vec{3}},\vec{1},+1/3)$
	 & $(\vec{1},\vec{1},0)$
	     & 
%		 & \eqref{eq:L-B-dN-u-d-NN}
		 \\
 %%%%%
 $(d_{R} N_{R i}) (d_{R}) (u_{R}) (N_{R j} N_{R k})$
 & $(\vec{3},\vec{1},-1/3)$
     & $(\overline{\vec{3}},\vec{1},-2/3)$
	 & $(\vec{1},\vec{1},0)$
	     & 
%		 & \eqref{eq:L-B-dN-d-u-NN}
		 \\
 %%%%%
 \hline
 $(u_{R} N_{R i}) (N_{R j})(N_{R k}) (d_{R} d_{R})$
 & $(\vec{3},\vec{1},+2/3)$
     & $(\vec{3},\vec{1},+2/3)$
	 & $(\overline{\vec{3}},\vec{1},-2/3)$
	     & $d_{R}d_{R}S_{2}^{\dagger} = 0$
%		 & \eqref{eq:L-B-uN-N-N-dd}
		 \\
 %%%%%
 $(u_{R} N_{R i}) (d_{R})(d_{R}) (N_{R j} N_{R k})$
 & $(\vec{3},\vec{1},+2/3)$
     & $(\overline{\vec{3}},\vec{1},+1/3)$
	 & $(\vec{1},\vec{1},0)$
	     & $\epsilon^{IJK}(d_{R I}d_{R J})=0$
%		 & \eqref{eq:L-B-uN-d-d-NN}
	     \\
%%%%%
 $(d_{R} d_{R})(u_{R})(N_{R i})(N_{R j} N_{R k})$
 & $(\overline{\vec{3}},\vec{1},-2/3)$
     & $(\vec{1},\vec{1},0)$
	 & $(\vec{1},\vec{1},0)$
	     & $d_{R}d_{R} S_{1}^{\dagger}=0$
%		 & \eqref{eq:L-B-dd-u-N-NN}
		 \\
%%%%%
 $(d_{R} d_{R})(N_{R i})(u_{R})(N_{R j} N_{R k})$
 & $(\overline{\vec{3}},\vec{1},-2/3)$
     & $(\overline{\vec{3}},\vec{1},-2/3)$
	 & $(\vec{1},\vec{1},0)$
	     & $d_{R}d_{R}S_{1}^{\dagger}=0$
%		 & \eqref{eq:L-B-dd-N-u-NN}
		 \\
 %%%%%
 \hline
 $(u_{R} N_{R i})(d_{R})(N_{R j})(d_{R} N_{R k})$
 & $(\vec{3},\vec{1},+2/3)$
     & $(\overline{\vec{3}},\vec{1},+1/3)$
	 & $(\vec{3},\vec{1},-1/3)$
	     & 
%		 & \eqref{eq:L-B-uN-d-N-dN}
	     \\
 %%%%%
 $(u_{R} N_{R i})(N_{R j})(d_{R})(d_{R} N_{R k})$
 & $(\vec{3},\vec{1},+2/3)$
     & $(\vec{3},\vec{1},+2/3)$
	 & $(\vec{3},\vec{1},-1/3)$
	     &
%		 & \eqref{eq:L-B-uN-N-d-dN}
	     \\
 %%%%%
 $(d_{R} N_{R i})(u_{R})(N_{R j})(d_{R} N_{R k})$
 & $(\vec{3},\vec{1},-1/3)$
     & $(\overline{\vec{3}},\vec{1},+1/3)$
	 & $(\vec{3},\vec{1},-1/3)$
	     & %$S_{1} \neq S_{2}$ 
%		 & \eqref{eq:L-B-dN-u-N-dN}
		 \\
 \hline \hline
\end{tabular}
\caption{Decompositions and the necessary mediator fields of the $d=9$ effective operator $u_{R} d_{R} d_{R} N_{R} N_{R} N_{R}$ with Topology B in Fig.~\ref{Fig:topology-d9}.
%
% Topology A and B altogether, there are 15 decompositions in total,
% which coincides with the result of 
% \texttt{Length[GenerateModelDiagrams[...]]}
% by \texttt{ModelGenerator}.
 }
 \label{Tab:decomposition-B}
\end{table}
%%%%%%%%%%%%%%%%%%%%%%%%%%%%%%%%%%%%%%%%%%%%%%%%%%%%%%%%%%%%%%%%%%%%%%

%%%%%%%%%%%%%%%%%%%%%%%%%%%%%%%%%%%%%%%%%%%%%%%%%%%%%%%%%%%%%%%%%%%%%%
\begin{table}[t]
\begin{tabular}{ccccc}
 \hline \hline
 Decomposition & $S_{1}/V_{1}$ & $S_{2}/V_{2}$ & $S_{3}/V_{3}$ & Comment %& Eq.
		 \\
 \hline
 $(QQ)(d_{R} N_{R i})(N_{R j}N_{R k})$
 & $S_{1}(\overline{\vec{3}},\vec{1},+1/3)$
     & $S_{2}(\vec{3},\vec{1},-1/3)$
	 & $S_{3}(\vec{1},\vec{1},0)$
	     & %$S_{1}$ must be an $SU(2)_{L}$ singlet
%		 & \eqref{eq:L-A-QQ-dN-NN}
     \\
 %%%%%
 $(Q d_{R})(Q N_{R i})(N_{R j} N_{R k})$
 & $V_{1}(\overline{\vec{3}},\vec{2},-1/6)$
     & $V_{2}(\vec{3},\vec{2},+1/6)$
	 & $S_{3}(\vec{1},\vec{1},0)$
	     &
%		 & \eqref{eq:L-A-Qd-QN-NN}
     \\
 %%%%%
 $(Q N_{R i})(Q N_{R j})(d_{R} N_{R k})$
 & $V_{1}(\vec{3},\vec{2},+1/6)$
     & $V_{2}(\vec{3},\vec{2},+1/6)$
	 & $S_{3}(\vec{3},\vec{1},-1/3)$
	     & $V_{1} = V_{2}$
		 %Can be $V_{1} = V_{2}$,
		% but $\mu_{V_{1}V_{2}S_{3}}$ does not vanish!
%		  & \eqref{eq:L-A-QN-QN-dN}
     \\
 \hline \hline
\end{tabular}
\caption{Decompositions of the $QQd_{R}N_{R}N_{R}N_{R}$ operator with
 Topology A in Fig.~\ref{Fig:topology-d9}.}
\label{Tab:QQdNNN-decomposition-A}
\end{table}
%%%%%%%%%%%%%%%%%%%%%%%%%%%%%%%%%%%%%%%%%%%%%%%%%%%%%%%%%%%%%%%%%%%%%%

%%%%%%%%%%%%%%%%%%%%%%%%%%%%%%%%%%%%%%%%%%%%%%%%%%%%%%%%%%%%%%%%%%%%%%
\begin{table}[t]
\begin{tabular}{ccccc}
 \hline \hline
 Decomposition & $S_{1}/V_{1}$ & $F$ & $S_{2}/V_{2}$ & Comment %& Eq.
		 \\
 \hline
 $(QQ)(N_{R i})(N_{R j})(d_{R}N_{R k})$
 & $S_{1}(\overline{\vec{3}},\vec{1},+1/3)$
     & $(\overline{\vec{3}},\vec{1},+1/3)$
	 & $S_{2}(\vec{3},\vec{1},-1/3)$
	     &
%		 & \eqref{eq:L-B-QQ-N-N-dN}
 \\
 %%%%%
 $(QQ)(d_{R})(N_{R i})(N_{R j} N_{R k})$
 & $S_{1}(\overline{\vec{3}},\vec{1},+1/3)$
     & $(\vec{1},\vec{1},0)$
	 & $S_{2}(\vec{1},\vec{1},0)$
	     & $F$ can be Majorana
%		 & \eqref{eq:L-B-QQ-d-N-NN}
 \\
 %%%%%
 $(QQ)(N_{R i})(d_{R})(N_{R j} N_{R k})$
 & $S_{1}(\overline{\vec{3}},\vec{1},+1/3)$
     & $(\overline{\vec{3}},\vec{1},+1/3)$
	 & $S_{2}(\vec{1},\vec{1},0)$
	     &
%		 & \eqref{eq:L-B-QQ-N-d-NN}
 \\
 %%%%%
 $(d_{R} N_{R i})(Q)(Q)(N_{R j} N_{R k})$
 & $S_{1}(\vec{3},\vec{1},-1/3)$
     & $(\overline{\vec{3}},\vec{2},-1/6)$
	 & $S_{2}(\vec{1},\vec{1},0)$
	     &
%		 & \eqref{eq:L-B-dN-Q-Q-NN}
 \\
 %%%%%%%%%%
 \hline
 $(Q d_{R})(N_{R i})(N_{R j})(Q N_{R k})$
 & $V_{1}(\overline{\vec{3}},\vec{2},-1/6)$
     & $(\overline{\vec{3}},\vec{2},-1/6)$
	 & $V_{2}(\vec{3},\vec{2},+1/6)$
	     &
%		 & \eqref{eq:L-B-Qd-N-N-QN}
 \\
 %%%%%
 $(Q d_{R})(Q)(N_{R i})(N_{R j} N_{R k})$
 & $V_{1}(\overline{\vec{3}},\vec{2},-1/6)$
     & $(\vec{1},\vec{1},0)$
	 & $S_{2}(\vec{1},\vec{1},0)$
	     & $F$ can be Majorana
%		 & \eqref{eq:L-B-Qd-Q-N-NN}
 \\
 %%%%%
 $(Q d_{R})(N_{R i})(Q)(N_{R j} N_{R k})$
 & $V_{1}(\overline{\vec{3}},\vec{2},-1/6)$
     & $(\overline{\vec{3}},\vec{2},-1/6)$
	 & $S_{2}(\vec{1},\vec{1},0)$
	     &
%		 & \eqref{eq:L-B-Qd-N-Q-NN}
 \\
 %%%%%
 $(Q N_{R i})(Q)(d_{R})(N_{R j} N_{R k})$
 & $V_{1}(\vec{3},\vec{2},+1/6)$
     & $(\overline{\vec{3}},\vec{1},+1/3)$
	 & $S_{2}(\vec{1},\vec{1},0)$
	     &
%		 & \eqref{eq:L-B-QN-Q-d-NN}
 \\
 %%%%%
 $(Q N_{R i})(d_{R})(Q)(N_{R j} N_{R k})$
 & $V_{1}(\vec{3},\vec{2},+1/6)$
     & $(\overline{\vec{3}},\vec{2},-1/6)$
	 & $S_{2}(\vec{1},\vec{1},0)$
	     &
%		 & \eqref{eq:L-B-QN-d-Q-NN}
 \\
 %%%%%%%%%%
 \hline
 $(Q N_{R i})(d_{R})(N_{R j})(Q N_{R k})$
 & $V_{1}(\vec{3},\vec{2},+1/6)$
     & $(\overline{\vec{3}},\vec{2},-1/6)$
	 & $V_{2}(\vec{3},\vec{2},+1/6)$
	     & $V_{1}=V_{2}$
%		 & \eqref{eq:L-B-QN-d-N-QN}
 \\
 %%%%%
 $(Q N_{R i})(Q)(N_{R j})(d_{R} N_{R k})$
 & $V_{1}(\vec{3},\vec{2},+1/6)$
     & $(\overline{\vec{3}},\vec{1},+1/3)$
	 & $S_{2}(\vec{3},\vec{1},-1/3)$
	     &
%		 & \eqref{eq:L-B-QN-Q-N-dN}
 \\
 %%%%%
 $(Q N_{R i})(N_{R j})(Q)(d_{R} N_{R k})$
 & $V_{1}(\vec{3},\vec{2},+1/6)$
     & $(\vec{3},\vec{2},+1/6)$
	 & $S_{2}(\vec{3},\vec{1},-1/3)$
	     &
%		 & \eqref{eq:L-B-QN-N-Q-dN}
 \\
 \hline \hline
\end{tabular}
\caption{Decompositions of the $QQd_{R}N_{R}N_{R}N_{R}$ operator 
with Topology B in Fig.~\ref{Fig:topology-d9}.}
\label{Tab:QQdNNN-decomposition-B}
\end{table}
%%%%%%%%%%%%%%%%%%%%%%%%%%%%%%%%%%%%%%%%%%%%%%%%%%%%%%%%%%%%%%%%%%%%%%

%%%%%%%%%%%%%%%%%%%%%%%%%%%%%%%%%%%%%%%%%%%%%%%%%%%%%%%%%%%%%%%%%%%%%%
%%%%%%%%%%%%%%%%%%%%%%%%%%%%%%%%%%%%%%%%%%%%%%%%%%%%%%%%%%%%%%%%%%%%%%

%%%%%%%%%%%%%%%%%%%%%%%%%%%%%%%%%%%%%%%%%%%%%%%%%%%%%%%%%%%%%%%%%%%%%%
\begin{table}[t]
\begin{tabular}{cccc}
 \hline \hline
 Decomposition & $S_{1}/V_{1}/F_{1}$ & $S_{2}/V_{2}/F_{2}$ & Comment %& Eq.
		 \\
 \hline
 $[\partial] (d_{R} d_{R})(u_{R})(\overline{N_{R}} a)$ 
 & $S_{1}(\overline{\vec{3}},\vec{1},-2/3)$
     & $F_{2}(\vec{1},\vec{1},0)$
	 & $d_{R}d_{R}S_{1}^{\dagger}=0$
%	     & \eqref{eq:dd-u-Na}
     \\
 %%%%%
 $[\partial] (d_{R} \overline{N_{R}}) (u_{R})(d_{R} a)$
 & $V_{1}(\vec{3},\vec{1},-1/3)$
     & $F_{2}(\vec{3},\vec{1},-1/3)$
	 &
%	     & \eqref{eq:dN-u-da}
 \\
 \hline
 %%%%%
 $[\partial] (u_{R} d_{R})(d_{R})(\overline{N_{R}} a)$
 & $S_{1}(\overline{\vec{3}},\vec{1},+1/3)$
     & $F_{2} (\vec{1},\vec{1},0)$
	 & $F_{2}$ can be Majorana
%		 & \eqref{eq:ud-d-Na}
     \\
 %%%%%
 $[\partial] (u_{R} \overline{N_{R}})(d_{R})(d_{R} a)$
 & $V_{1}(\vec{3},\vec{1},+2/3)$
     & $F_{2}(\vec{3},\vec{1},-1/3)$
	 &
%	     & \eqref{eq:uN-d-da}
     \\
 %%%%%
 $[\partial] (u_{R} a) (d_{R})(d_{R} \overline{N_{R}})$
 & $F_{1}(\vec{3},\vec{1},+2/3)$
     & $V_{2}(\vec{3},\vec{1},-1/3)$
	 &
%	     & \eqref{eq:ua-d-dN}
     \\
 %%%%%
 \hline
 $[\partial] (u_{R} d_{R})(\overline{N_{R}})(d_{R} a)$
 & $S_{1}(\overline{\vec{3}},\vec{1},+1/3)$
     & $F_{2}(\vec{3},\vec{1},-1/3)$
	 &
%	     & \eqref{eq:ud-N-da}
     \\
 %%%%%
 $[\partial] (u_{R} a) (\overline{N_{R}}) (d_{R} d_{R})$
 &  $F_{1}(\vec{3},\vec{1},+2/3)$
     & $S_{2}(\overline{\vec{3}},\vec{1},-2/3)$
	 & $d_{R}d_{R}S_{2}^{\dagger}=0$ 
%	     & \eqref{eq:ua-N-dd}
     \\
 %%%%%
 \hline
 $[\partial] (u_{R} d_{R}) (a) (d_{R} \overline{N_{R}})$
 & $S_{1}(\overline{\vec{3}},\vec{1},+1/3)$
     & $V_{2}(\vec{3},\vec{1},-1/3)$
	 &
%	     & \eqref{eq:ud-a-dN}
     \\
 %%%%%
 $[\partial] (u_{R} \overline{N_{R}}) (a) (d_{R} d_{R})$
 & $V_{1}(\vec{3},\vec{1},+2/3)$
     & $S_{2}(\overline{\vec{3}},\vec{1},-2/3)$
	 & $d_{R}d_{R}S_{2}^{\dagger}=0$
%	     & \eqref{eq:uN-a-dd}
	     \\
 \hline \hline
\end{tabular}
\caption{Decompositions of $u_{R} d_{R} d_{R} \overline{N_{R}} a$, 
 which
 result in the $d=8$ effective operator with a derivative operator,
 $\partial a u_{R} d_{R} d_{R} \overline{N_{R}} $.
 The operator $u_{R} d_{R} d_{R} \overline{N_{R}} Z'$ with a vector
 boson is decomposed with the same mediators.}
\label{Tab:uddNa-decomposition}
\end{table}
%%%%%%%%%%%%%%%%%%%%%%%%%%%%%%%%%%%%%%%%%%%%%%%%%%%%%%%%%%%%%%%%%%%%%%
%%%%%%%%%%%%%%%%%%%%%%%%%%%%%%%%%%%%%%%%%%%%%%%%%%%%%%%%%%%%%%%%%%%%%%
\begin{table}[t]
\begin{tabular}{cccc}
 \hline \hline
 Decomposition & $S_{1}/V_{1}/F_{1}$ & $S_{2}/V_{2}/F_{2}$ & Comment %& Eq.
		 \\
 \hline
 $[\partial] (Q Q) (d_{R}) (\overline{N_{R}} a)$
 & $S_{1}(\overline{\vec{3}},\vec{1},+1/3)$
     & $F_{2}(\vec{1},\vec{1},0)$
	 & $F$ can be Majorana
%	     & \eqref{eq:L-QQ-d-Nca}
		 \\
 %%%%%
 $[\partial] (Q \overline{N_{R}}) (d_{R}) (Q a)$
 & $S_{1}(\vec{3},\vec{2},+1/6)$
     & $F_{2}(\vec{3},\vec{2},+1/6)$
	 &
%	     & \eqref{eq:L-QNc-d-Qa}
		 \\
 %%%%%
 \hline
 $[\partial] (Q d_{R}) (Q) (\overline{N_{R}} a)$
 & $V_{1}(\overline{\vec{3}},\vec{2},-1/6)$
     & $F_{2}(\vec{1},\vec{1},0)$
	 & $F$ can be Majorana
%	     & \eqref{eq:L-Qd-Q-Nca}
		 \\
 %%%%%
 $[\partial] (Q \overline{N_{R}}) (Q) (d_{R} a)$
 & $S_{1}(\vec{3},\vec{2},+1/6)$
     & $F_{2}(\vec{3},\vec{1},-1/3)$
	 &
%	     & \eqref{eq:L-QNc-Q-da}
		 \\
 %%%%%
 $[\partial] (Q a) (Q) (d_{R} \overline{N_{R}})$
 & $F_{1}(\vec{3},\vec{2},+1/6)$
     & $V_{2}(\vec{3},\vec{1},-1/3)$
	 &
%	     & \eqref{eq:L-Qa-Q-dNc}
		 \\
 %%%%%
 \hline
 $[\partial] (Q d_{R}) (\overline{N_{R}}) (Q a)$
 & $V_{1}(\overline{\vec{3}},\vec{2},-1/6)$
     & $F_{2}(\vec{3},\vec{2},+1/6)$
	 &
%	     & \eqref{eq:L-Qd-Nc-Qa}
		 \\
 %%%%%
 $[\partial] (d_{R} a) (\overline{N_{R}}) (Q Q)$
 & $F_{1}(\vec{3},\vec{1},-1/3)$
     & $S_{2}(\overline{\vec{3}},\vec{1},+1/3)$
	 &
%	     & \eqref{eq:L-da-Nc-QQ}
		 \\
 %%%%%
 \hline
 $[\partial] (Q d_{R}) (a) (Q \overline{N_{R}})$
 & $V_{1}(\overline{\vec{3}},\vec{2},-1/6)$
     & $S_{2}(\vec{3},\vec{2},+1/6)$
	 &
%	     & \eqref{eq:L-Qd-a-QNc}
		 \\
 %%%%%
 $[\partial] (d_{R} \overline{N_{R}}) (a) (Q Q)$
 & $V_{1}(\vec{3},\vec{1},-1/3)$
     & $S_{2}(\overline{\vec{3}},\vec{1},+1/3)$
	 &
%	     & \eqref{eq:L-dNc-a-QQ}
		 \\
 \hline \hline
\end{tabular}
 \caption{Decompositions of $Q Q d_{R} \overline{N_{R}} a$,
 which
 result in the $d=8$ effective operator with a derivative operator,
 $\partial a Q Q d_{R} \overline{N_{R}}$.
 The operator $QQ d_{R} \overline{N_{R}} Z'$ is decomposed with the same
 mediators.}
 \label{Tab:QQdNa-decomposition}
\end{table}
%%%%%%%%%%%%%%%%%%%%%%%%%%%%%%%%%%%%%%%%%%%%%%%%%%%%%%%%%%%%%%%%%%%%%%

%%%%%%%%%%%%%%%%%%%%%%%%%%%%%%%%%%%%%%%%%%%%%%%%%%%%%%%%%%%%%%%%%%%%%%
\begin{table}[t]
\begin{tabular}{cccc}
 \hline \hline
 Decomposition & $S_{1}/F_{1}$ & $S_{2}/F_{2}$ & Comment %& Eq.
		 \\
 \hline
 $(d_{R} d_{R})(u_{R})(N_{R} \phi)$ 
 & $S_{1}(\overline{\vec{3}},\vec{1},-2/3)$
     & $F_{2}(\vec{1},\vec{1},0)$
	 & $d_{R} d_{R} S_{1}^{\dagger}=0$ 
%	     & \eqref{eq:L-dd-u-Np}
     \\
 %%%%%
 $(d_{R} N_{R}) (u_{R})(d_{R} \phi)$
 & $S_{1}(\vec{3},\vec{1},-1/3)$
     & $F_{2}(\vec{3},\vec{1},-1/3)$
	 &
%	     & \eqref{eq:L-dN-u-dp}
 \\
 \hline
 %%%%%
 $(u_{R} d_{R})(d_{R})(N_{R} \phi)$
 & $S_{1}(\overline{\vec{3}},\vec{1},+1/3)$
     & $F_{2}(\vec{1},\vec{1},0)$
	 & $F_{2}$ can be Majorana
%		 & \eqref{eq:L-ud-d-Np}
     \\
 %%%%%
 $(u_{R} N_{R})(d_{R})(d_{R} \phi)$
 & $S_{1}(\vec{3},\vec{1},+2/3)$
     & $F_{2}(\vec{3},\vec{1},-1/3)$
	 &
%	     & \eqref{eq:L-uN-d-dp}
     \\
 %%%%%
 $(u_{R} \phi) (d_{R})(d_{R} N_{R})$
 & $F_{1}(\vec{3},\vec{1},+2/3)$
     & $S_{2}(\vec{3},\vec{1},-1/3)$
	 &
%	     & \eqref{eq:L-up-d-dN}
     \\
 %%%%%
 \hline
 $(u_{R} d_{R})(N_{R})(d_{R} \phi)$
 & $S_{1}(\overline{\vec{3}},\vec{1},+1/3)$
     & $F_{2}(\vec{3},\vec{1},-1/3)$
	 &
%	     & \eqref{eq:L-ud-N-dp}
     \\
 %%%%%
 $(u_{R} \phi) (N_{R}) (d_{R} d_{R})$
 & $F_{1}(\vec{3},\vec{1},+2/3)$
     & $S_{2}(\overline{\vec{3}},\vec{1},-2/3)$
	 & $d_{R}d_{R}S_{2}^{\dagger}=0$
%	     & \eqref{eq:L-up-N-dd}
     \\
 %%%%%
 \hline
 $(u_{R} d_{R}) (\phi) (d_{R} N_{R})$
 & $S_{1}(\overline{\vec{3}},\vec{1},+1/3)$
     & $S_{2}(\vec{3},\vec{1},-1/3)$
	 & $S_{2}^{\dagger} \neq S_{1}$ to avoid $d=6$ 
	     $u_{R} d_{R} d_{R} N_{R}$
%	     & \eqref{eq:L-ud-p-dN}
     \\
 %%%%%
 $(u_{R} N_{R}) (\phi) (d_{R} d_{R})$
 & $S_{1}(\vec{3},\vec{1},+2/3)$
     & $S_{2}(\overline{\vec{3}},\vec{1},-2/3)$
	 & $d_{R}d_{R}S_{2}^{\dagger}=0$ 
%	     & \eqref{eq:L-uN-p-dd}
	     \\
 \hline \hline
\end{tabular}
\caption{Decompositions of $u_{R} d_{R} d_{R} N_{R} \phi$.}
\label{Tab:uddNp-decomposition}
\end{table}
%%%%%%%%%%%%%%%%%%%%%%%%%%%%%%%%%%%%%%%%%%%%%%%%%%%%%%%%%%%%%%%%%%%%%%
%%%%%%%%%%%%%%%%%%%%%%%%%%%%%%%%%%%%%%%%%%%%%%%%%%%%%%%%%%%%%%%%%%%%%%
\begin{table}[t]
\begin{tabular}{cccc}
 \hline \hline
 Decomposition & $S_{1}/V_{1}/F_{1}$ & $S_{2}/V_{2}/F_{2}$ 
	 & Comment %& Eq.
		 \\
 \hline
 $(Q d_{R})(Q)(N_{R} \phi)$
 & $V_{1}(\overline{\vec{3}},\vec{2},-1/6)$
     & $F_{2}(\vec{1},\vec{1},0)$
	 & $F_{2}$ can be Majorana
%	     & \eqref{eq:L-Qd-Q-Np}
 \\
 %%%%%
 $(Q N_{R}) (Q) (d_{R} \phi)$
 & $V_{1}(\vec{3},\vec{2},+1/6)$
     & $F_{2}(\vec{3},\vec{1},-1/3)$
	 &
%	     & \eqref{eq:L-QN-Q-dp}
 \\
 %%%%%
 $(Q \phi) (Q) (d_{R} N_{R})$
 & $ F_{1}(\vec{3},\vec{2},+1/6)$
     & $S_{2}(\vec{3},\vec{1},-1/3)$
	 &
%	     & \eqref{eq:L-Qp-Q-dN}
 \\
 %%%%%
 \hline
 $(QQ) (d_{R}) (N_{R} \phi)$
 & $S_{1}(\overline{\vec{3}},\vec{1},+1/3)$
     & $F_{2}(\vec{1},\vec{1},0)$
	 & $F_{2}$ can be Majorana
%	     & \eqref{eq:L-QQ-d-Np}
 \\
 %%%%%
 $(QN_{R})(d_{R})(Q\phi)$
 & $V_{1}(\vec{3},\vec{2},+1/6)$
     & $F_{2}(\vec{3},\vec{2},+1/6)$
	 &
%	     & \eqref{eq:L-QN-d-Qp}
 \\
 %%%%%
 \hline
 $(QQ)(N_{R})(d_{R} \phi)$
 & $S_{1}(\overline{\vec{3}},\vec{1},+1/3)$
     & $F_{2}(\vec{3},\vec{1},-1/3)$
	 &
%	     & \eqref{eq:L-QQ-N-dp}
 \\
 %%%%%
 $(Qd_{R})(N_{R})(Q \phi)$
 & $V_{1}(\overline{\vec{3}},\vec{2},-1/6)$
     & $F_{2}(\vec{3},\vec{2},+1/6)$
	 &
%	     & \eqref{eq:L-Qd-N-Qp}
 \\
 %%%%%
 \hline
 $(QQ)(\phi)(d_{R} N_{R})$
 & $S_{1}(\overline{\vec{3}},\vec{1},+1/3)$
     & $S_{2}(\vec{3},\vec{1},-1/3)$
	 & $S_{2}^{\dagger} \neq S_{1}$ to avoid $d=6$ $QQd_{R}N_{R}$
%	     & \eqref{eq:L-QQ-p-dN}
 \\
 %%%%%
 $(Qd_{R})(\phi)(Q N_{R})$
 & $V_{1}(\overline{\vec{3}},\vec{2},-1/6)$
     & $V_{2} (\vec{3},\vec{2},+1/6)$
	 & $V_{2}^{\dagger} \neq V_{1}$ to avoid $d=6$ $QQd_{R} N_{R}$
%	     & \eqref{eq:L-Qd-p-QN}
 \\
 \hline \hline
\end{tabular}
\caption{Decompositions of $QQ d_{R} N_{R} \phi$.}
\label{Tab:QQdNp-decomposition}
\end{table}
%%%%%%%%%%%%%%%%%%%%%%%%%%%%%%%%%%%%%%%%%%%%%%%%%%%%%%%%%%%%%%%%%%%%%%

%%%%%%%%%%%%%%%%%%%%%%%%%%%%%%%%%%%%%%%%%%%%%%%%%%%%%%%%%%%%%%%%%%%%%%
\section{A ``black box theorem'' for $\Delta(L)=3$ operators}
\label{app2}

In the introduction, we have argued that with only the particle
content of the standard model it is highly unlikely that invisible
neutron decay can exist with measurable rates, without also having
nucleon decays with charged leptons in the final state with at least
comparable rates. In this appendix, we will discuss the basis of this
claim in more detail.

As we have mentioned in the introduction, it is possible to forbid all
$d=9$ operators by some symmetry. However, as we will now proceed
to show, this is not possible at $d=10$. 
Concretely, we will show that if the $d=12$ SMEFT operator of
Eq.~\eqref{eq:W12} exists, than also the $d=10$ $\Delta(L)=3$
operator, ${\cal O}_{d=10} \propto d^cd^cd^cLLLH$, must have a
non-zero Wilson coefficient. This proof is modeled on the
``black box theorem'' for neutrinoless double beta decay
\cite{Schechter:1981bd,Hirsch:2006yk}. This ``theorem'' is a
proof by contradiction. Given two particular operators, 
one assumes that one of them exists, while the other is forbidden
by some symmetry. One then proceeds to show that no such
symmetry can exist and that either both operators or none of
them will be present in the Lagrangian.

For the formulation of this proof, in addition to the two operators of
interest, ${\cal O}_{d=10}$ and ${\cal O}_{d=12}^{\text{W}}$
(see Eq.~\eqref{eq:W12}), we will need the following standard model Yukawa
interactions for up and down quarks:
\begin{equation}\label{eq:yuk}
  {\cal L}^{\rm Yuk} = Y_d \overline{Q_L}\cdot H d_R
                   +  Y_u \overline{Q_L}\cdot H^{\dagger}u_R
                   \hskip1mm {\rm + h.c.}
\end{equation}  
Then, assume there exists a global symmetry, which we denote as
$G_\eta$, under which the SM fields transform as:
\begin{eqnarray}\label{eq:tr}
  H \stackrel{G_\eta}{\longrightarrow} \eta_H \cdot H & & \\ \nonumber
  L  \stackrel{G_\eta}{\longrightarrow} \eta_L \cdot L, & & 
  Q_L \stackrel{G_\eta}{\longrightarrow} \eta_Q  \cdot Q_L \\ \nonumber
  u_R \stackrel{G_\eta}{\longrightarrow} \eta_u  \cdot u_R,
  & & d_R \stackrel{G_\eta}{\longrightarrow} \eta_d  \cdot d_R
\end{eqnarray}  
The Yukawa interactions in Eq.~\eqref{eq:yuk} then give the
constraints:
\begin{equation}\label{eq:GYuk}
\eta_{Q}^\dagger \eta_d \eta_H =1 \hskip10mm {\rm and} \hskip10mm
\eta_{Q}^\dagger \eta_u \eta_H^\dagger =1 .
\end{equation}  
Now, assume ${\cal O}_{d=12}^{\text{W}}$ is allowed by this symmetry,
while ${\cal O}_{d=10}$ is forbidden. This leads to the constraints:
\begin{equation}\label{eq:GOp}
  \eta({\cal O}_{d=12}^{\text{W}})  \hskip2mm \to \hskip2mm
  \eta_u^\dagger (\eta_d^\dagger)^2 (\eta_L)^3 (\eta_H)^3  =1 \hskip10mm {\rm and}
  \hskip10mm
  \eta({\cal O}_{d=10}) \hskip2mm \to \hskip2mm
   (\eta_d^\dagger)^3 (\eta_L)^3 \eta_H  \ne 1 
\end{equation}  
Using Eq.~\eqref{eq:GYuk} one finds $\eta_u=\eta_d \eta_H^2$. Inserting
this relation into Eq.~\eqref{eq:GOp} yields:
\begin{equation}\label{eq:GOp2}
  \eta({\cal O}_{d=12}^{\text{W}}) = \eta({\cal O}_{d=10}),
\end{equation}  
which contradicts our assumption. Thus, either both operators,
${\cal O}_{d=12}^{\text{W}}$ and ${\cal O}_{d=10}$ are allowed or both
are forbidden. 

This symmetry argument makes it clear, why invisible neutron decay can
not exist in SMEFT without nucleon decays with charged leptons in the
final state, in principle. However, this kind of ``proof'' based on
symmetries strictly speaking only guarantees that the Wilson
coefficients of the operators of interest are both non-zero -- it does
not require them to be equal. We stress, however, even if $c_{10}$
were zero at a high scale, RGE running would generate a non-zero
$c_{10}$ at the low scale -- precisely because $c_{10}$ can not be
protected by a symmetry.

Since the RGEs for these high-dimensional operators are not known, we
can not quantify the effect. Instead, we will ask the question, how
much fine-tuning one would need to accept in order to suppress decays
generated from ${\cal O}_{d=10}$ to an acceptable rate.  To make a
quantitative statement, we therefore compare the predicted half-lives
for the ${\cal O}_{d=12}^{\text{W}}$ and ${\cal O}_{d=10}$ operators with the
corresponding experimental sensitivities.  For ${\cal O}_{d=12}^{\text{W}}$,
Eq.~\eqref{eq:wdth12} gives our estimate.  For ${\cal O}_{d=10}$, we
estimate the partial width for the decay $n \to e^- \pi^+ \nu\nu$ as:
\begin{eqnarray}\label{eq:wdth10}
  \Gamma \sim \frac{W_0(\pi)^2 m_N^{7} \langle H^{0} \rangle^2}
         {589824 \pi^5 \Lambda^{12}}
  & \Rightarrow & \frac{1}{3.3 \cdot 10^{23}\hskip1mm {\rm yr}}
  \Big(\frac{13.4\hskip1mm {\rm TeV}}{\Lambda}\Big)^{12} .
\end{eqnarray}  
Both, Eq.~\eqref{eq:wdth12} and Eq.~\eqref{eq:wdth10} assume the
Wilson coefficient to be equal to one, and the widths scale as
the Wilson coefficients squared.

To the best of our knowledge, there is no dedicated search for $n \to
e^- \pi^+ \nu\nu$. However, we can use the Super-Kamiokande search for
$n \to e^+ \pi^-$ \cite{Super-Kamiokande:2017gev} to derive a
conservative limit on $n \to e^- \pi^+ \nu\nu$. Based on a statistics
of $0.316$ megaton$\cdot$years and a background estimate of $0.41 \pm 0.13$
events, Super-Kamiokande gives a limit of $5.3 \cdot 10^{33}$ ys
\cite{Super-Kamiokande:2017gev} on $n \to e^+ \pi^-$. In Fig.~18
Super-Kamiokande provides the distribution of $e + \pi $ events as a
function of the total momentum. The experimental collaboration uses
the MonteCarlo expectation for the signal events of this distribution
to cut backgrounds. Since the final state $n \to e^- \pi^+ \nu\nu$
involves missing momentum, we can use the results of Fig.~18 to give
a very rough, but also very conservative limit on $n \to e^- \pi^+
\nu\nu$ by simply assuming that the number of events for $n \to e^-
\pi^+ \nu\nu$ does not exceed the number of {\it all} background
events. This simple-minded but conservative procedure leads to an
estimate for the lower limit on the half-live of $n \to e^- \pi^+
\nu\nu$ of roughly $(1.5-3.5) \times 10^{31}$ ys.\footnote{%
The sum of all events in Fig.~18 \cite{Super-Kamiokande:2017gev}
is roughly $\sim 300$. For a half-life as given in Eq.~\eqref{eq:wdth10}
one would expect roughly $10^8$ events.}

For the ${\cal O}_{d=10}$ to generate a decay rate obeying this bound,
while using $\Lambda=13.4$ TeV, the Wilson coefficient $c_{10}$
must be smaller than $c_{10} \lsim 10^{-4}$. Recall, that in the
estimate, Eq.~\eqref{eq:wdth12}, we have used $c_{12}^{\text{W}}=1$. Thus,
at least four orders of fine-tuning in the Wilson coefficient
$c_{10}$ would be necessary, to avoid the constraints coming from
searches with charged lepton final states.

%%%%%%%%%%%%%%%%%%%%%%%%%%%%%%%%%%%%%%%%%%%%%%%%%%%%%%%%%%%%%%%%%%%%%%
\bibliography{N-inv-d9}
\bibliographystyle{h-physrev5}

\end{document}